\documentclass[lettersize,journal]{IEEEtran}
\usepackage[utf8]{inputenc}
\usepackage[T1]{fontenc}
\usepackage{cite}
\usepackage{caption} 
\usepackage{amsmath,amssymb,amsfonts}
\usepackage{algorithmic}
\usepackage{pdfpages}
\usepackage{graphicx}
\usepackage{booktabs}
\usepackage{textcomp}
\usepackage{xcolor}
\usepackage{tablefootnote}
\usepackage[misc]{ifsym}
\usepackage{subfig}
\usepackage{adjustbox}
\usepackage{subcaption}
\usepackage{amsmath,amssymb,amsfonts}
\usepackage{graphicx}
\usepackage{amsthm}
\usepackage{multirow}
\usepackage{bibentry}
\usepackage{listings} 
\usepackage{float}
\usepackage{color}
\usepackage{url}
\usepackage[ruled,linesnumbered]{algorithm2e}
\usepackage{makecell}
\usepackage{booktabs}
\usepackage{amsthm}  

\begin{document}

\title{Communication-Efficient Collaborative LLM Inference over LEO Satellite Networks
}
\author{Songge~Zhang,~\IEEEmembership{Graduate Student Member,~IEEE,}
        Wen~Wu,~\IEEEmembership{Senior Member,~IEEE,} Liang~Li,~\IEEEmembership{Member,~IEEE,} Ye~Wang,~\IEEEmembership{Member,~IEEE,} 
        and~Xuemin~(Sherman)~Shen,~\IEEEmembership{Fellow,~IEEE}
\thanks{
Songge Zhang is with the School of Electronic and Computer Engineering, Peking University Shenzhen Graduate School, Shenzhen, 518055, China, and also with the Frontier Research Center, Pengcheng Laboratory, Shenzhen, 518055, China (email: zhangsongge@stu.pku.edu.cn);

Wen Wu, Liang Li, and Ye Wang are with the Pengcheng Laboratory, Shenzhen, 518055, China (email: \{wuw02, lil03, wangy02\}@pcl.ac.cn).


Xuemin (Sherman) Shen is with the Department of Electrical and Computer Engineering, University of Waterloo, Waterloo, N2L 3G1, Canada (email: sshen@uwaterloo.ca).}

}

\maketitle
\begingroup\renewcommand\thefootnote{${\textrm{\Letter}}$}

\endgroup
\begin{abstract}
Low Earth orbit (LEO) satellites play an essential role in intelligent Earth observation by leveraging artificial intelligence models. However, limited onboard memory and excessive inference delay prevent the practical deployment of large language models (LLMs) on a single satellite.
In this paper, we propose a communication-efficient collaborative LLM inference scheme for LEO satellite networks. Specifically, the entire LLM is split into multiple sub-models, with each deployed on a satellite, thereby enabling collaborative LLM inference via exchanging intermediate activations between satellites. The proposed scheme also leverages the pipeline parallelism mechanism that overlaps sub-model inference with intermediate activation transmission, thereby reducing LLM inference delay. An adaptive activation compression scheme is designed to mitigate cumulative errors from multi-stage model splitting while preserving inference accuracy. Furthermore, we formulate the LLM inference delay minimization problem by jointly optimizing model splitting and compression ratios under onboard memory and inference accuracy constraints.
The problem is transformed into a shortest-path search problem over a directed acyclic graph that edge weights explicitly quantify the inference delay induced by model splitting and compression strategies, which is solved via a modified $\mathrm{A}^{\star}$-based search algorithm. Extensive simulation results indicate that the proposed solution can reduce inference delay by up to 42\% and communication overhead by up to 71\% compared to state-of-the-art benchmarks, while maintaining the inference accuracy loss of less than 1\%.
\end{abstract}

\section{Introduction}
Low Earth orbit (LEO) satellites have become an essential network infrastructure for the sixth generation (6G) networks, providing communication, sensing, and computing services~\cite{shen2, Backgroud1}.
An increasing number of LEO satellites have been launched into space to provide ubiquitous connectivity. For example, Starlink has launched over 4,000 satellites, and other mega-constellations, such as OneWeb and Telesat, are also rapidly expanding~\cite{ZhangHongKe}. 
Recent advances in satellite hardware have endowed LEO satellites with substantial on-board computing capabilities, \textcolor{black}{e.g., graphics processing units (GPUs), enabling on-orbit deployment of artificial intelligence (AI) models.} The onboard computing facilitates real-time data processing and rapid responses to downstream tasks, such as Earth observation, disaster response, navigation, and environmental monitoring~\cite{Background3}.
The on-orbit paradigm shift significantly reduces communication overhead by processing raw sensing data on board, while transmitting only the computed results to ground stations~\cite{Background2}. 
\textcolor{black}{Deploying large language models (LLMs), such as vision transformers (ViT), on satellites offers good adaptability to the downstream tasks}~\cite{WUWC}.

LLM inference on LEO satellites is constrained by limited onboard memory and computational resources, since LLMs typically contain billions of parameters~\cite{NLP,ShiwenMao,XiaopingZhang2}. The massive footprint may exceed onboard memory capacity and render deployment infeasible, or incur long inference delay when deployable. Researchers have developed model compression techniques, including structured weight sparsification~\cite{Distillation}, pruning~\cite{Pruning}, and low-bit quantization~\cite{Quantization}. These methods substantially reduce model size, enabling execution within the tight resource envelope of satellites. However, such parameter reduction methods may inevitably lead to performance degradation, potentially compromising the inference accuracy.

Model splitting is a promising technique for deploying LLM sub-models across multiple satellites to address the limitations of model compression~\cite{shen3}. This strategy alleviates the resource constraints on individual satellites and enables collaborative inference. Nevertheless, existing studies typically adopt a single-split scheme, which is insufficient for satellite networks requiring distributed satellite execution. In multi-split scenarios, the heterogeneity of satellite computing resources can result in significant idle time and waiting time, increasing the overall delay of the serial inference~\cite{YuhaoChen,Songge1}. Moreover, inter-satellite transmissions become more frequent, resulting in increased communication overhead.
Therefore, \textcolor{black}{it is essential to explore a collaborative and communication-efficient scheme in heterogeneous satellite environments.}

\begin{figure}[t]
    \centering
    \includegraphics[width=3.3in]{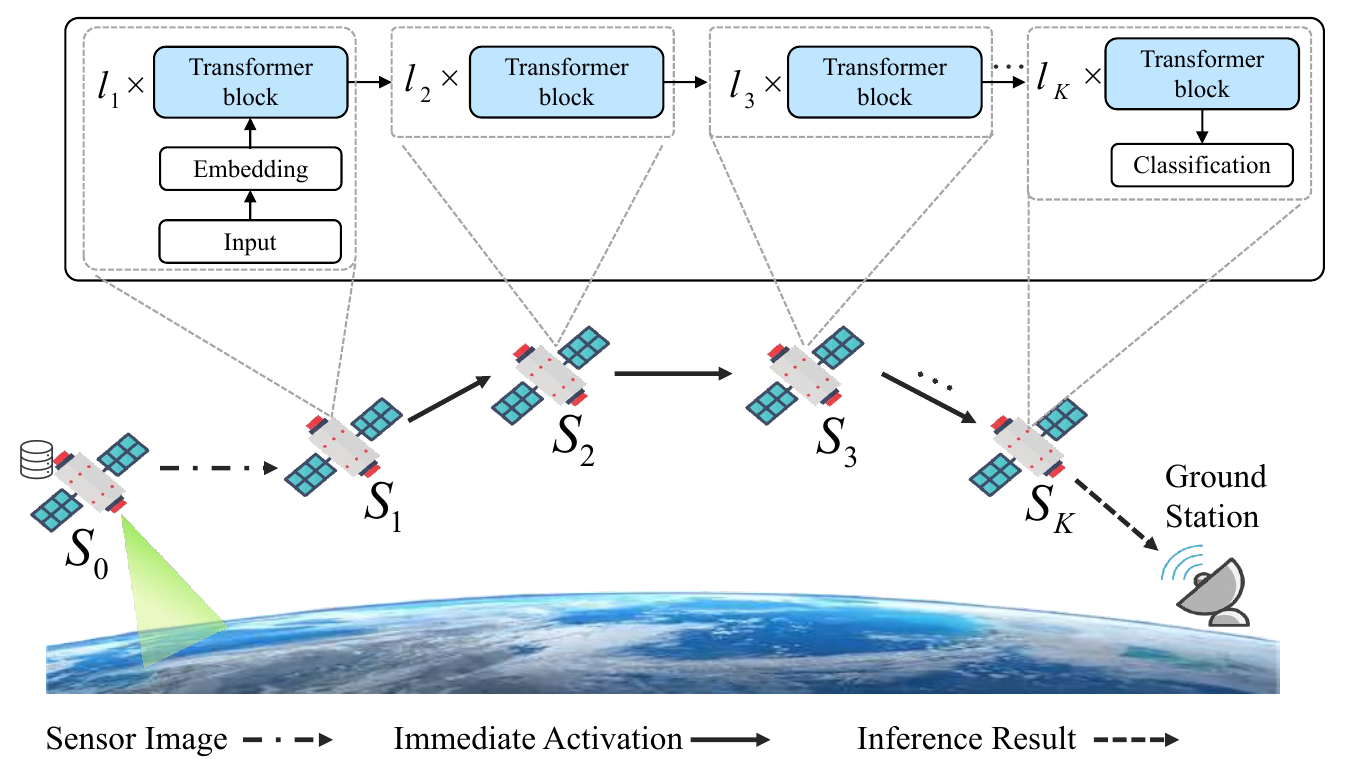}
    \caption{An overview of the collaborative LLM inference scheme in satellite networks.}
    \label{fig:Scenario}
    \vspace{-1.0em}
\end{figure}

In this paper, we \textit{first} propose a collaborative inference scheme across multiple LEO satellites to accelerate the LLM inference. 
Specifically, we split an LLM into multiple sub-models and deploy them across different satellites, thereby meeting the memory constraints of individual satellites, as illustrated in Fig.~\ref{fig:Scenario}. The multiple satellites are executed collaboratively, with intermediate activation sequentially transmitted via inter-satellite links to improve inference efficiency. \textcolor{black}{We reduce the collaborative inference delay through pipeline parallelism, which allows each satellite to overlap the activation communication with the computation of its sub-model. \textit{Second}, an adaptive compression scheme is proposed to reduce inter-satellite communication overhead incurred by transmitting intermediate activations. A learnable Gumbel-mask module is trained end-to-end with a straight-through estimator and sparsity regularization to select informative activations, producing near-binary masks that sparsify the transmitted activations.}
The selected features are then quantized into low-bit formats and further compressed through entropy-based coding.

Furthermore, we formulate a minimization problem to optimize the LLM splitting strategy and compression ratios subject to constraints on onboard memory and inference accuracy.
\textcolor{black}{The objective is the total inference delay, and the formulation is a mixed integer nonlinear programming problem.}
To solve it, the minimization problem is reformulated as a shortest-path search over a directed acyclic graph (DAG), where each path represents a valid layer assignment to satellites, and the path cost corresponds to the total inference delay. 
An iterative $\mathrm{A}^{\star}$-based algorithm is then designed, where the inner loop solves for the optimal compression ratios, and the outer loop searches for the optimal LLM splitting strategy.
\textcolor{black}{Experimental results demonstrate that the proposed scheme reduces inference delay by up to 42\%, and communication overhead by up to 71.7\% compared to state-of-the-art benchmarks, while maintaining inference accuracy loss within 1\%.}
The main contributions of this paper are summarized as follows:
\begin{enumerate}
    \item We propose a collaborative LLM inference scheme across multiple LEO satellites through model splitting and pipeline parallelism;

    \item We propose an adaptive and learnable compression scheme, which can reduce inter-satellite communication overhead while preserving inference accuracy;
    
    \item We formulate a delay minimization problem to jointly optimize the model splitting strategy and activation compression ratios;

    \item We design a directed acyclic graph-based solution and propose a modified $A^\star$ algorithm to obtain the optimal decision variables through an outer–inner search.
\end{enumerate}

\textcolor{black}{
The remainder of this paper is organized as follows. Section~II reviews related work. Section~III presents the proposed collaborative LLM inference scheme and activation compression scheme. Section~IV formulates the end-to-end inference delay minimization problem. Section~V describes the proposed solution. Section VI presents the experimental results, and Section VII concludes the paper.}
Table~\ref{tab:notations} summarizes the main symbols.


\begin{table}[t]
\footnotesize
\centering
\caption{List of Important Notations}
\label{tab:notations}
\begin{tabular}{l|l}
\toprule
\textbf{Notation} & \textbf{Description} \\
\hline
$L$ & Total number of layers in the LLM \\
$K$ & Number of computing satellites\\
$l_k$ & Layers assigned to satellite $k$, $\sum_{k=1}^K l_k = L$ \\
$f_k$ & Computational capacity of satellite $k$ (FLOPs/s) \\
$C_k(l_k)$ & FLOPs required at satellite $k$ with $l_k$ layers \\
$T_k^{\text{comp}}$ & Computation delay at satellite $k$ \\
$q_k $ & Compression ratio at satellite $k$ \\
$S_k$ & Feature size at output of satellite $k$  \\
$T_k^{\text{comm}}$ & Communication delay from satellite $k$ to $k{+}1$ \\
$T_k^{\text{eff}}$ & Steady-state delay at satellite $k$ considering overlap \\
$T_{\text{total}}$ & Total inference delay  \\
$\theta$ & Auxiliary variable \\
$M_k(l_k)$ & Memory usage at satellite $k$ under $l_k$ layers \\
$M_k^{\text{max}}$ & Maximum memory capacity of satellite $k$ \\
$\text{Acc}_k(q_k)$ & Accuracy at satellite $k$ under compression $q_k$ \\
$\text{Acc}_{\min}$ & Minimum required inference accuracy \\
$T$ & Number of input samples \\
$y^t_K$ & Predicted result for input $d^t$ \\
$S$ & Sequence length of token \\
$D$ & Hidden dimension of transformer features \\
$P$ & Number of pixels per image \\
$E$ & Number of output classes \\
$r_{\text{sat}}$ & Inter-satellite data rate  \\
$r_{\text{gs}}$ & Ground-to-satellite data rate \\
\bottomrule
\end{tabular}
\end{table}

\section{Related Work}
To meet the growing demands for low-latency Earth observation and autonomous decision-making in space, on-orbit computing has emerged as a promising paradigm that shifts AI inference from ground stations to satellites themselves. 
Typically, satellite systems primarily acted as passive data collectors, transmitting raw data to the ground station for processing~\cite{Vincentpoor}. However, the ground-based processing paradigm suffers from severe downlink bottlenecks, especially when dealing with high-resolution imagery or continuous video streams. 
To alleviate the communication burden, recent efforts have focused on integrating lightweight onboard filtering techniques that prioritize or discard data based on relevance.  For instance, Giuffrida \emph{et al}. demonstrated that edge-deployed convolutional neural networks (CNNs) could effectively identify low-utility images in real time~\cite{Back1}. Similarly, Cratere \emph{et al} implemented FPGA-accelerated CNNs onboard satellites to filter irrelevant image content before transmission~\cite{Back2}. These approaches ensured that only mission-critical data is sent to the ground, thereby significantly improving bandwidth utilization.
\textcolor{black}{The advancement of AI-optimized embedded hardware, including central processing units (CPUs), GPUs, and neural processing units (NPUs), has enabled the direct deployment of inference models onboard satellites.} 
These hardware systems support real-time deep neural network inference without requiring a ground server.
Nonetheless, deploying a large AI model in orbit remains challenging due to limitations in energy and memory. To address these constraints, researchers have developed model compression techniques, including structured weight sparsification~\cite{Distillation}, pruning~\cite{Pruning}, and low-bit quantization~\cite{Quantization}, which substantially reduce model size, enabling onboard execution within the resource-limited satellites. Zhang \emph{et al}.  proposed an object knowledge distilled joint detection and tracking framework,  to perform object detection in onboard satellite videos~\cite{Back4}. Sun \emph{et al}. proposed a pruning method that integrates graph topology information with model pruning to optimize the graph neural network structure and communication overhead in distributed satellite communications~\cite{Back5}. Since model compression inevitably reduces parameter capacity and may degrade performance, collaborative inference across multiple nodes emerges as a promising solution to support LLM inference on onboard satellites without compromising accuracy.


Collaborative inference has emerged as an effective paradigm for accelerating AI inference tasks by leveraging the computational resources of multiple distributed nodes~\cite{OJVT}. 
Existing collaborative inference methods can be broadly categorized into data partitioning and model splitting. Among these, model splitting has gained significant attention due to its ability to decouple the AI model and distribute sub-models across different nodes. In the most common two-node setting, the model is divided into two parts: the front-end layers are deployed on the user-side device (e.g., satellite or edge terminal), while the remaining layers reside on the server-side or ground station~\cite{ShangguangWang, Chenxu}. The split paradigm allows for early feature extraction on resource-constrained nodes, followed by high-capacity processing downstream.
Sun \emph{et al}. proposed a two-part model splitting scheme, which dynamically adjusts the splitting point and allocates resources in real time to meet user-specific AI inference tasks~\cite{Back6}.
Li \emph{et al}. introduced a delay-aware DNN inference framework that improves throughput in mobile edge computing environments by jointly optimizing model splitting and parallel multi-thread execution~\cite{Back7}.
In more complex multi-node scenarios, the model can be divided into multiple parts, each assigned to a different node in the system~\cite{INFOCOM}.
Cao \emph{et al}. proposed a multi-layer partitioned multi-access edge computing system, which aims to meet users' computation latency requirements through the joint optimization of communication and computing resource allocation~\cite{Back8}. Xie \emph{et al}. investigated the design of multi-layer edge computing architectures and the co-scheduling of heterogeneous edge computing resources~\cite{Back9}.
To further accelerate multi-node collaboration, Narayanan \emph{et al}. introduced a pipeline mechanism that enables model splitting and parallel processing across multiple GPUs, thereby improving overall training throughput and reducing per-machine latency~\cite{Pipedream}. Wang \emph{et al}. considered the pipeline mechanism in wireless networks and formulated the model multi-splitting problem to optimize distributed algorithms, thereby reducing communication overhead and improving overall inference efficiency~\cite{Xinyu}.
However, model splitting introduces additional communication overhead, which becomes more pronounced in multi-segment scenarios. To address this, some studies have explored compression mechanisms to reduce the transmission of intermediate activations~\cite{LiWu}. For example, Zheng \emph{et al}. proposed a randomized Top-E sparsification method that selects the top E activation elements for transmission~\cite{TopK1}. These works enable the deployment of LLMs by splitting the model into multiple segments across distributed nodes and improve communication efficiency by compressing intermediate activations exchanged between model segments. Different from the existing works, the proposed scheme considers the collaborative inference pipeline parallelism for LEO satellite networks, and an adaptive compression strategy is tailored for multi-part partitioning scenarios, which effectively reduces communication overhead while preserving model inference accuracy. In addition, we formulate a joint optimization problem of layer partitioning and compression ratio, and develop a graph-based algorithm to efficiently solve it.

\section{Proposed Collaborative LLM Inference Scheme}

\subsection{System Model}
In this paper, a distributed LEO satellite network consisting of heterogeneous satellites is considered, in which on-orbit LLM inference for intelligent Earth observation tasks is collaboratively performed.
As illustrated in Fig.~\ref{fig:Scenario}, the network comprises a sensing satellite $S_0$ equipped with optical sensors, and a set of computing satellites $\mathcal{S}_C = \{S_1, S_2, \dots, S_K\}$ interconnected via inter-satellite links. The sensing satellite $S_0$ is responsible for capturing high-resolution remote sensing images and initiating the LLM inference process. These raw images are then delivered to the computing satellite chain for distributed processing.
Each computing satellite $S_k \in \mathcal{S}_C$ is equipped with onboard processors and memory resources for handling a sub-model of the LLM.

Due to the limited computational capacity and memory of individual satellites, deploying an entire LLM on a single satellite for inference is infeasible. To address this, we adopt a collaborative LLM inference scheme in which the model is partitioned into multiple components and deployed across multiple satellites, with intermediate activations exchanged to enable end-to-end inference. However, the frequent inter-satellite transmission of activations incurs substantial communication overhead. Therefore, we propose an activation compression scheme, which is detailed in the following two subsections.

\begin{figure}[t]
    \centering
    \includegraphics[width=2.8in]{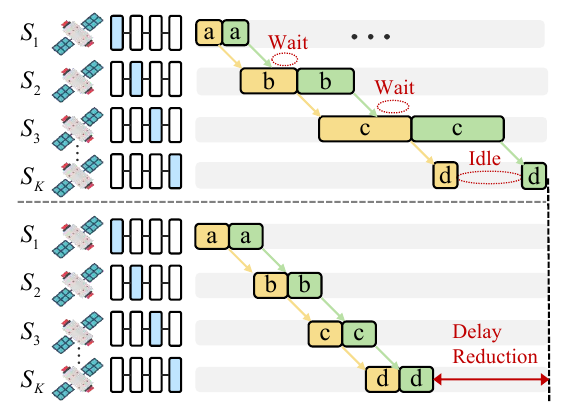}
    \caption{Parallel computing process for model splitting.}

    \label{fig:pipeline}
    \vspace{-1.0em}
\end{figure}

\subsection{Collaborative LLM Inference Scheme}
A collaborative LLM inference scheme for satellite networks is proposed, consisting of model splitting, collaborative inference, and pipeline parallelism, as detailed below.

\subsubsection{Model Splitting}
In the proposed scheme, when a satellite is assigned a heavy computation workload or constrained memory resources, the target LLM can be split into multiple sub-models and deployed across several satellites. As illustrated in Fig.~1, the model is divided into $K$ sub-models and each sub-model is assigned to a computing satellite in a chain of $L$ satellites, where satellite $S_k$ is allocated $l_k$ LLM layers.
Due to the inherently sequential nature of collaborative inference, unbalanced layer assignments can lead to idle and waiting times. As illustrated in Fig.~2, consider two batches (yellow and green) with $K=4$ satellites ($S_1$–$S_K$) hosting sub-models $a$–$d$. In the first case, the total time for the yellow batch to execute $a$, transfer to $b$, and complete $b$ exceeds the time for the green batch to finish $a$ and transfer to $b$, causing the green batch to wait on $S_2$, which means that $S_1$ is underloaded while $S_2$ is overloaded. In the second case, computing $c$ and transferring to $d$ for the yellow batch is much faster than the arrival of the green batch at $S_K$, leaving $S_K$ idle while $S_3$ is overloaded. To address such imbalances, the model splitting point needs to be optimized and the splitting algorithm will be presented in Section~IV.

\subsubsection{Collaborative Inference}
Based on the model splitting point, a complete LLM is divided into $K$ sub-models, denoted as $\mathbf{w} = \{w_1, w_2, \cdots, w_K\}$. The overall inference process is described as follows.
We consider an inference task on a dataset $\mathcal{D}$ containing $D$ images. The dataset is divided into $T$ mini-batches $d^1, d^2, \dots, d^T$, with $\sum_{t=1}^{T} |d^t| = D$. For each mini-batch,  the inference begins at the first sub-model $w_1$ deployed on satellite $S_1$, expressed as
$s_1^t = f_1(d^t, w_1^t),$
where $f_1(\cdot)$ denotes the mapping function from the input to the intermediate output of the first sub-model, consisting of embedding layers, Transformer blocks, and output layers specific to the LLM. The intermediate activation $s_1^t$ is then transmitted via the inter-satellite link to satellite $S_2$, serving as input to the next sub-model, which can be 
$s_2^t = f_2(s_1^t, w_2^t).$
The inference process continues through the remaining sub-models, where each sub-model $w_k$ on satellite $S_k$ takes as input the activation from the previous stage, and is formulated as
$s_k^t = f_k(s_{k-1}^t, w_k^t), \quad \text{for } k = 2, 3, \dots, K.$
The final output from the last sub-model is the predicted result and can be expressed as
$y_K^t = f_K(s_{K-1}^t, w_K^t).$
After all batches are processed, the final results $\{y_K^t \mid t = 1, 2, \dots, T\}$ are aggregated at satellite $S_K$ and transmitted to the ground station via the satellite-to-ground link.

\subsubsection{Pipeline Parallelism}
In conventional systems, batch processing occurs sequentially due to tight coupling between computation and communication phases. Each computational stage remains idle until its required inputs are fully transmitted, creating deterministic execution dependencies that serialize pipeline workflows. For instance, the computation of $f_2$ must wait until all $s_1^t$ are received, and the transmission of $s_2^t$ must wait until $f_2$ has finished processing. As a result, the inference delay grows rapidly with increasing batch size.
To address this, we propose a pipeline-based parallel execution scheme that allows satellites to compute and communicate concurrently. Specifically, while satellite $S_1$ is transmitting $s_1^t$, it can immediately begin computing $s_1^{t+1}$. Each satellite $S_k$ can start computing $s_k^{t+1}$ while transmitting $s_k^t$. This concurrent execution can be implemented using multiple threading techniques to enable parallel processing of computation and communication, ensuring non-blocking processing.
Following the proposed scheme, all satellites can remain active at all times, significantly improving the resource efficiency and accelerating the inference process.

\subsection{Activation Compression Scheme}
An activation compression scheme is proposed that combines learnable sparsification for adaptively selecting informative activations with subsequent quantization and entropy coding to further reduce the transmission, as detailed below.
\subsubsection{Gumbel Mask-based Sparsification}
A lightweight and fully differentiable sparse compression module, primarily guided by Gumbel masking, is proposed to selectively retain the most informative activations while discarding less relevant ones.
Given the input activation tensor $\mathbf{X} \in \mathbb{R}^{N \times S \times D}$, the goal is to generate a binary mask $\mathbf{M} \in \{0, 1\}^{N \times S \times D}$ that selects informative features. To preserve differentiability, the mask is approximated using a Gumbel-Sigmoid sampling mechanism.
Each element of the mask is parameterized by a trainable logit $\alpha_{ij}$ and perturbed with Gumbel noise to simulate sampling. The noise term $G_{ij}$ is drawn from the standard Gumbel distribution using inverse transform sampling, computed as $G_{ij} = -\log(-\log(U_{ij}))$, where $U_{ij} \sim \text{Uniform}(0,1)$. This sampled noise is then added to $\alpha_{ij}$ before applying temperature scaling and a sigmoid function to obtain a continuous approximation of the binary mask.
The perturbed logits are scaled by a temperature parameter $\tau > 0$ and then passed through a sigmoid function, resulting in a continuous soft mask, and can be expressed as
\begin{equation}
\hat{M}_{ij} = \sigma\left( \frac{\alpha_{ij} + G_{ij}}{\tau} \right).
\end{equation}
A smaller $\tau$ yields mask values that are closer to 0 or 1, effectively approximating discrete binary decisions, while a larger $\tau$ results in smoother, more probabilistic outputs between 0 and 1.

To obtain discrete binary masks while maintaining gradient flow, the Straight-Through Estimator (STE) is applied. During the forward pass, the mask is binarized as
\begin{equation}
M_{ij} = \mathbb{I}(\hat{M}_{ij} > 0.5),
\end{equation}
and during the backward pass, the gradient is approximated as $\nabla M_{ij} \approx \nabla \hat{M}_{ij}$.

The sparse activation is then computed as
\begin{equation}
\tilde{\mathbf{X}} = \mathbf{M} \odot \mathbf{X} + (1 - \mathbf{M}) \odot \operatorname{stopgrad}(\mathbf{X}),
\end{equation}
where $\odot$ denotes element-wise multiplication and $\operatorname{stopgrad}(\cdot)$ preserves the forward value while blocking gradients from the deactivated features.

To promote sparsity in the mask, a regularization term is introduced to constrain the expected number of activated elements. The sparsity loss is defined as
\begin{equation}
\mathcal{L}_{\text{sparse}} = \lambda  \frac{1}{SD} \sum_{i=1}^{S} \sum_{j=1}^{D} \sigma(\alpha_{ij}),
\end{equation}
where $\lambda$ controls the sparsity level, and this term is jointly optimized with the task loss.

To facilitate training, the temperature $\tau$ is gradually decreased during training using an annealing schedule:
\begin{equation}
\tau(t) = \max\left( \tau_{\min}, \tau_0 \cdot \left(1 - \frac{t}{T} \right) \right),
\end{equation}
where $t$ is the current epoch, $T$ is the total number of epochs, and $\tau_0$, $\tau_{\min}$ are initial and minimal temperatures. Gradually decreasing $\tau$ allows the mask to transition from soft exploration to hard selection as training progresses.

\subsubsection{Quantization}

Only the non-zero elements retained by the binary mask $\mathbf{M}$ are subject to quantization. Given a target bit-width $b$, the quantization step size is computed as
\begin{equation}
\Delta = \frac{x_{\max} - x_{\min}}{2^{b-1} - 1},
\end{equation}
where $x_{\min}$ and $x_{\max}$ denote the minimum and maximum absolute values of the active elements in the current batch.

Each non-zero element $\tilde{x}_{ijk}$ is first quantized by mapping it to an integer code using
$
q_{ijk} = \text{sign}(\tilde{x}_{ijk})  \left\lfloor \frac{|\tilde{x}_{ijk}| - x_{\min}}{\Delta} + 0.5 \right\rfloor,
$
and then dequantized back to the real domain using
$
\hat{x}_{ijk} = \text{sign}(\tilde{x}_{ijk})  \left( x_{\min} + q_{ijk}  \Delta \right).
$
Since the quantization process involves non-differentiable rounding operations, the straight-through estimator (STE) is employed to preserve gradient flow, in which the gradient of the dequantized value is approximated by that of the original input.
To mitigate the influence of outliers, the quantization range $[x_{\min}, x_{\max}]$ is dynamically estimated from each mini-batch. If no elements are active in a batch, quantization is skipped, and the sparse tensor is directly forwarded to the next stage.

\subsubsection{Entropy-Guided coding}
To evaluate compressibility, the expected bit length is approximated using the entropy of the quantized value distribution:
\begin{equation}
H(\mathcal{S}) = -\sum_{v \in \mathcal{V}} p(v) \log_2 p(v),
\end{equation}
where $\mathcal{S}$ denotes the set of quantized non-zero values, $\mathcal{V}$ is the set of unique values, and $p(v)$ is the empirical frequency of symbol $v$. 
Based on this entropy estimate, the uncompressed bit length is calculated as $L_{\text{raw}} = |\mathcal{S}|  b$, where $b$ is the fixed quantization bit-width. The corresponding estimated compressed bit length using entropy coding (e.g., Huffman coding) is given by $L_{\text{huff}} \approx |\mathcal{S}|  H(\mathcal{S})$.
The entropy value reflects the amount of redundancy in the quantized representation. A lower entropy indicates that fewer bits are required on average to encode the data, while a higher entropy corresponds to a more uniform distribution and thus lower compressibility.

\section{Problem Formulation}

\subsection{Computation and Communication Model}
\subsubsection{Computing Delay}
To quantify the computational delay incurred by each satellite during distributed inference, both the computational workloads and the memory capacities of the satellites are taken into account.
Let $C_k$ denote the total number of floating-point operations (FLOPs) required to execute the sub-model $w_k$ assigned to satellite $S_k$. Let $f_k$ represent the computing capability of satellite $S_k$, measured in FLOPs per second. 
Accordingly, the computation delay of satellite $S_k$ for a single mini-batch is given by
$T_k^{\text{comp}} = \frac{C_k}{f_k},$
where $T_k^{\text{comp}}$ denotes the time required to complete the inference of one batch through the sub-model $w_k$ on satellite $S_k$.

\subsubsection{Communication Delay}
In collaborative satellite inference, communication comprises three distinct phases: (i) input image upload from the sensing satellite to the initial computing satellite, (ii) intermediate activation transfer between successive computing satellites via inter-satellite links, and (iii) final result download from the last computing satellite to ground stations. The corresponding data sizes are denoted as $S_{\text{input}}$, $S_k^{\text{act}}$, and $S_{\text{out}}$, respectively. For phase (i), the input image size is given by $S_{\text{input}} = N P$, where $N$ is the batch size and $P$ is the number of pixels per image. For phase (ii), the intermediate activation size at stage $k$ is modeled as $S_k^{\text{act}} = N S D$, where $D$ is determined by the LLM architecture and $S$ is the sequence length used in inference~\cite{Songge2}. For phase (iii), the final output comprises classification logits or predicted labels over $E$ categories, leading to an output size of $S_{\text{out}} = N E$.

The communication delay at each phase depends on both the data volume and the available link bandwidth. Let $r_{\text{gs}}$ denote the satellite-to-ground data rate, and $r_{\text{sat}}$ denote the inter-satellite data rate. Accordingly, the delay for transmitting the input image from the ground station to the first satellite is $T_0^{\text{comm}} = S_{\text{input}} / r_{\text{gs}}$, the delay for transferring intermediate activations between satellites is $T_k^{\text{comm}} = S_k^{\text{act}} / r_{\text{sat}}$, and the delay for sending the final output back to the ground is $T_K^{\text{comm}} = S_{\text{out}} / r_{\text{gs}}$.

\subsection{LLM Inference Delay Model}
The inference process across $K$ pipeline stages, each executed on a computing satellite, consists of two distinct phases: a \textit{startup phase} and a \textit{steady-state phase}:
\subsubsection{Startup Delay}
The startup delay refers to the time required for the first input sample to traverse all $K$ stages sequentially, which is computed as the sum of computation and communication delays across all stages, as given by
\begin{equation}
T_{\text{startup}} = \sum_{k=1}^{K} \left( T_k^{\text{comp}} + T_k^{\text{comm}} \right).
\end{equation}

\subsubsection{Steady-State Delay}
Once the startup is completed, each subsequent input can be processed in a steady-state stage. The delay for the steady-satege stage is determined by the slowest satellite, which becomes the bottleneck. Hence, the steady-state stage delay is given by
$T_{\text{steady}} = \max_k T_k^{\text{eff}}.$
Based on the pipeline, computation and communication can overlap. For satellite $k$, the overlapping duration is
\begin{equation}
T_k^{\text{overlap}} = \min \left( T_k^{\text{comp}}, T_{k-1}^{\text{comm}} \right),
\end{equation}
which captures the maximal parallel time between computing and receiving data.
Accordingly, the effective delay is
\begin{equation}
T_k^{\text{eff}} = T_k^{\text{comp}} + T_k^{\text{comm}} - T_k^{\text{overlap}}.
\end{equation}

\subsubsection{Total Inference Delay}
For a batch of $d$ input samples, the total end-to-end inference delay consists of three components. First, there is the initial transmission delay, denoted as $T_0^{\text{comm}}$, which represents the time required to send the data from the ground station to the first satellite. Second, the startup delay, $T_{\text{startup}}$, accounts for the time needed to fill the pipeline, during which the first sample sequentially passes through all $K$ stages. Third, after the pipeline is filled, the remaining $B - 1$ samples are processed in a pipelined manner, incurring a steady-state delay $T_{\text{steady}}$ that is determined by the slowest stage in the pipeline.
Thus, the total inference delay is
\begin{equation}
T_{\text{total}} = T_0^{\text{comm}} + T_{\text{startup}} + (B - 1) \max_k T_k^{\text{eff}}.
\end{equation}

\subsection{Accuracy and Memory Model}
Activation compression may cause approximation errors and degrade the LLM inference accuracy. 
Directly modeling the accuracy as a function of the full compression vector $\mathbf{q}=[q_1,\ldots,q_{K-1}]$ is intractable due to the high-dimensional combinatorial space. 
Instead, the relationship between accuracy and compression is calibrated along the setting $q_1=\cdots=q_{K-1}=q$. 
Specifically, an offline sweep over compression ratios $q\in\mathcal{Q}$ is performed and the corresponding end-to-end accuracy is measured on a calibration set, thereby obtaining empirical pairs ${(q,a(q))}$.
A monotone regression function $\widehat{\mathrm{Acc}}(q)$ is then fitted to characterize the relationship between compression and accuracy. 
In~\eqref{eq:accuracy_constraint}, the per-stage accuracy function is defined using the same for all stages:
\begin{equation}
\mathrm{Acc}_k(q_k) \triangleq \widehat{\mathrm{Acc}}(q_k), \quad \forall k\in\{1,\ldots,K-1\}.
\end{equation}
Accordingly, the constraint $\mathrm{Acc}_k(q_k)\ge \mathrm{Acc}_{\min}$ serves as a sufficient and conservative condition to ensure that compression on each inter-satellite transmission meets the required accuracy threshold.

The onboard memory consumption is similarly modeled as a function of the assigned layers. Specifically, an offline profiling procedure is conducted to measure the memory usage under different layer allocations $l_k$, and a fitted function $M_k(l_k)$ is obtained for each satellite $k$. The resulting memory model is incorporated into the optimization via the constraint $M_k(l_k) \leq M_k^{\max},\ \forall k\in\{1,\ldots,K\}$, so that the layer assignment is constrained by the onboard memory budget of every satellite.

\subsection{Problem formulation}
An optimization problem is formulated to minimize the total inference delay for the proposed collaborative LLM inference scheme. The objective is to jointly optimize the model-splitting strategy and the inter-satellite activation compression ratios by assigning the $L$ model layers to $K$ computing satellites, subject to onboard memory and inference-accuracy constraints.

\subsubsection{Optimization Variables}

Let $l_k \in \mathbb{Z}^+$ denote the number of layers assigned to satellite $k$, satisfying the total layer constraint
$\sum_{k=1}^{K} l_k = L.$
The variable $l_k$ determines the computational workload at satellite $k$. Denote $C_k(l_k)$ as the total number of FLOPs required for the assigned layers, and $f_k$ as the computational capacity of satellite $k$, then the corresponding computation delay is
$T_k^{\text{comp}} = \frac{C_k(l_k)}{f_k}.$

To reduce communication overhead between adjacent satellites, we introduce a new variable $q_k \in (0, 1]$, representing the compression ratio applied to the output feature map at satellite $k$. A smaller $q_k$ indicates higher compression. Assuming the uncompressed feature size is $S_k$, the communication delay from satellite $k$ to $k+1$ is given by
$T_k^{\text{comm}} = \frac{q_k  S_k}{r_{\text{sat}}}.$

\subsubsection{Optimization Problem}
The final optimization problem is formulated as follows:
\begin{subequations}\label{Problem1}
\begin{align}
\min_{\{l_k\}, \{q_k\}} \quad & \sum_{k=1}^{K} \left( \frac{C_k(l_k)}{f_k} + \frac{q_k S_k}{r_{\text{sat}}} \right) + (B-1) \max_k T_k^{\text{eff}}(l_k, q_k) \\
\text{s.t.} \quad & \sum_{k=1}^{K} l_k = L,\quad l_k \in \mathbb{Z}^+, \label{eq:layer_constraint} \\
& M_k(l_k) \leq M_k^{\text{max}},\quad \forall k \in \{1,\dots,K\}, \label{eq:memory_constraint} \\
& q_k \in [0, 1],\quad \forall k \in \{1,\dots,K-1\}, \label{eq:compression_constraint} \\
& \text{Acc}_k(q_k) \geq \text{Acc}_{\min},\quad \forall k \in \{1,\dots,K-1\}. \label{eq:accuracy_constraint}
\end{align}
\end{subequations}

The constraint in~\eqref{eq:layer_constraint} ensures that the total number of layers allocated across all satellites equals the model depth \( L \).  
Constraint~\eqref{eq:memory_constraint} ensures the memory requirement \( M_k(l_k) \) on each satellite does not exceed its maximum onboard memory \( M_k^{\text{max}} \).  
Constraint~\eqref{eq:compression_constraint} constrains the compression ratios $q_k$ to lie in $[0,1]$, where $q_k = 1$ indicates no compression and smaller values correspond to more aggressive compression.  
Constraint~\eqref{eq:accuracy_constraint} imposes per-stage accuracy constraints, where \( \text{Acc}_k(q_k) \) represents the accuracy function with respect to compression ratio at stage \( k \), and \( \text{Acc}_{\min} \) is the acceptable lower bound.
The steady-state delay at stage $k$, denoted by $T_k^{\text{eff}}(l_k,q_k)$,  is given by
\begin{equation}
T_k^{\text{eff}}(l_k, q_k) = \frac{C_k(l_k)}{f_k} + \frac{q_k  S_k}{r_{\text{sat}}} - \min\left( \frac{C_k(l_k)}{f_k}, \frac{q_{k-1}  S_{k-1}}{r_{\text{sat}}} \right).
\end{equation}

The above optimization problem is non-trivial to solve due to the following three challenges. Firstly, it is a mixed-integer nonlinear programming (MINLP) problem involving both discrete variables \( \{l_k\} \in \mathbb{Z}^+ \) and continuous variables \( \{q_k\} \in [0,1] \), which are strongly coupled in the effective delay term \( \max_k T_k^{\text{eff}}(l_k, q_k) \), making the joint optimization more complex. Secondly, the maximum term \( \max_k T_k^{\text{eff}}(l_k, q_k) \) introduces non-smoothness, which prevents direct application of gradient-based or convex optimization techniques. Thirdly, the solution space expands exponentially with both the number of layers \( L \) and the number of computing stages \( K \). The combinatorial nature of layer partitioning under the constraint \( \sum_{k=1}^{K} l_k = L \), combined with the continuous search space of compression ratios \( \{q_k\} \), results in a high-dimensional search space and making exhaustive enumeration computationally prohibitive, especially for LLM.

To address the non-smooth $\max$ operator in the objective function, an auxiliary variable $\theta$ is introduced to represent the maximum effective stage delay across all satellites. The transformation allows the objective to be linearized by minimizing $\theta$, while the original $\max$ term is incorporated into the constraints. The reformulated optimization problem is expressed as
\begin{subequations}\label{RefinedProblem}
\begin{align}
\min_{\{l_k\}, \{q_k\}, \theta} \quad & \sum_{k=1}^{K} \left( \frac{C_k(l_k)}{f_k} + \frac{q_k S_k}{r_{\text{sat}}} \right) + (B - 1) \theta \\
\text{s.t.} \quad & T_k^{\text{eff}}(l_k, q_k) \leq \theta, \quad \forall k \in \{1, \dots, K\}, \label{eq:eff_constraint} \\
& \eqref{eq:layer_constraint}, \eqref{eq:memory_constraint}, \eqref{eq:compression_constraint}, \text{ and } \eqref{eq:accuracy_constraint}.
\end{align}
\end{subequations}

\section{Proposed Solution}

\subsection{Problem Reformulation}

To efficiently solve the layer partitioning problem defined in Eq.~\eqref{RefinedProblem}, the original problem is reformulated as a shortest-path problem over a directed acyclic graph (DAG). Each path in the graph corresponds to a valid assignment of model layers to satellites, and the objective is to find the path with the minimum total inference delay.

\subsubsection{Graph Construction}

A DAG $G = (V, E)$ is constructed to represent the space of layer-to-satellite assignment decisions. The source node is $(0, 0)$, indicating that no layers have been assigned yet, and the sink node is $(L, K)$, corresponding to a complete layer allocation over $K$ satellites.  
A directed edge $e = ((l, k), (l', k+1)) \in E$ exists if the $(k+1)$-th satellite is assigned layers $l+1$ through $l'$, i.e., $l_{k+1} = l' - l$, and the memory constraint
$M_{k+1}(l' - l) \leq M_{k+1}^{\text{max}}$
is satisfied.  
In addition to layer allocation, each edge implicitly determines a compression ratio \( q_{k+1} \in [0, 1] \) for the output features transmitted from satellite $k+1$ to $k+2$, which affects both communication delay and accuracy. The optimal value of \( q_{k+1} \) on each edge is determined through a per-edge local optimization process, which will be detailed in the next subsection.

\subsubsection{Edge Cost}

Each edge $e = ((l, k), (l', k+1))$ represents an assignment of $l_{k+1} = l' - l$ layers to satellite $S_{k+1}$ and carries two key attributes. First, the edge cost encodes the startup delay due to both computation and communication, which can be expressed as
\begin{equation}
C(e) = \frac{C_{k+1}(l' - l)}{f_{k+1}} + \frac{q_{k+1}^\star S_{k+1}}{r_{\text{sat}}},
\end{equation}
where \( q_{k+1}^\star \) denotes the compression ratio determined for this edge.  
Second, each edge also stores the stage-specific effective delay under pipelined execution:
\begin{align}\label{T_eff}
T^{\text{eff}}_{k+1}(l' - l) =\ & \frac{C_{k+1}(l' - l)}{f_{k+1}} + \frac{q_{k+1}^\star S_{k+1}}{r_{\text{sat}}} \nonumber \\
& - \min\left( \frac{C_{k+1}(l' - l)}{f_{k+1}}, \frac{q_{k}^\star S_{k}}{r_{\text{sat}}} \right),
\end{align}
where \( q_{k}^\star \) is the compression ratio from the preceding stage. Unlike edge costs, these effective delays are tracked separately for each edge and used to estimate path-level bottlenecks.

\subsubsection{Path Cost}

The layer allocation corresponds to a path $\mathcal{P}$ from $(0, 0)$ to $(L, K)$. The total inference delay associated with $\mathcal{P}$ is given by
\begin{equation}
C(\mathcal{P}) = \sum_{e \in \mathcal{P}} C(e) + (B - 1) \theta(\mathcal{P}),
\end{equation}
where $\theta(\mathcal{P}) = \max_{k} T_k^{\text{eff}}$ is the maximum effective delay among all stages in the path, corresponding to the steady-state bottleneck during pipelined execution.

\subsubsection{Problem Reformulation}
The original model-splitting optimization then reduces to a minimum-cost path search over the constructed DAG, which can be expressed as
\begin{equation}
\mathcal{P}^* = \arg\min_{\mathcal{P} \in \text{Paths}(0 \to L)} C(\mathcal{P}),
\end{equation}
subject to feasibility constraints such as memory usage, accuracy requirements, and complete layer coverage.

\subsection{Per-Edge Compression Optimization}
The compression ratio optimization is embedded as an inner-loop routine within the LLM splitting algorithm.
\subsubsection{Solution Strategy}
Given a current layer partitioning $\{l_k\}$ obtained from the graph-based reformulation, the compression ratios ${q_k}$ are next optimized to further reduce the total inference delay. The compression ratios directly influence the transmission time of intermediate features between adjacent satellites, and thereby affect both the stage-wise delay and the overall pipeline delay.

Each stage's effective delay includes a $\min$ operator coupling $q_k$ and $q_{k-1}$, which makes the optimization problem non-smooth and prevents closed-form solutions. To address this, the compression optimization is reformulated as a constrained numerical problem:
\begin{subequations}\label{eq:compression_subproblem}
\begin{align}
\min_{\{q_k\}, \theta} \quad & \sum_{k=1}^{K} \left( \frac{C_k(l_k)}{f_k} + \frac{q_k S_k}{r_{\text{sat}}} \right) + (B - 1)\theta \label{eq:compression_obj} \\
\text{s.t.} \quad & T_k^{\text{eff}}(l_k, q_k) \leq \theta, \quad \forall k \in \{1, \dots, K\}, \label{eq:compression_eff} \\
& q_k \in [0, 1], \quad \forall k \in \{1, \dots, K-1\}, \label{eq:compression_range} \\
& \text{Acc}_k(q_k) \geq \text{Acc}_{\min}, \quad \forall k \in \{1, \dots, K-1\}. \label{eq:compression_acc}
\end{align}
\end{subequations}
Due to the problem's low dimensionality, practical solutions include grid-based search or constrained optimization solvers such as sequential quadratic programming (SQP). In particular, we adopt a full-grid enumeration method with fixed resolution to exhaustively evaluate all feasible $\{q_k\}$ combinations and determine the optimal one that minimizes the objective in \eqref{eq:compression_obj}. The procedure is detailed in Alg.~\ref{alg:compression_search}.
The resulting optimal compression settings $\{q_k^*\}$ are then used to refine layer assignment in the next iteration. The alternating optimization procedure between layer partitioning and compression is described in the next section.

\subsubsection{Computational Complexity}
The total number of compression configurations to evaluate is $(N+1)^{K-1}$, where $K$ is the number of computing satellites and $N$ is the grid resolution for each compression ratio $q_k \in \{0, \frac{1}{N}, \dots, 1\}$. 
For each candidate setting, the algorithm evaluates $K$ stage-wise delay constraints and $K-1$ accuracy constraints, each assumed to take constant time. 
Therefore, the total computational complexity of Alg.~\ref{alg:compression_search} is 
$\mathcal{O}\left((N+1)^{K-1}  K\right).$

\begin{algorithm}[t]
\footnotesize
\caption{\textcolor{black}{Discrete Search for Compression Ratios}}
\label{alg:compression_search}
\begin{algorithmic}[1]
\REQUIRE Layer assignment $\{l_k\}$, accuracy functions $\text{Acc}_k(\cdot)$, grid resolution $N$
\STATE Initialize candidate set $\mathcal{Q} \leftarrow$ all $\{q_k\}$ with $q_k \in \{0, \frac{1}{N}, \dots, 1\}$
\STATE Best cost $\mathcal{C}^* \leftarrow +\infty$, Best setting $\{q_k^*\} \leftarrow \emptyset$
\FORALL{candidate $\{q_k\} \in \mathcal{Q}$}
    \STATE Compute $T_k^{\text{eff}}(l_k, q_k)$ and check constraint \eqref{eq:compression_eff}
    \STATE Evaluate accuracy constraint \eqref{eq:compression_acc}
    \IF{all constraints satisfied}
        \STATE Compute total delay cost $C_{\text{total}}$ using \eqref{eq:compression_obj}
        \IF{$C_{\text{total}} < \mathcal{C}^*$}
            \STATE Update $\mathcal{C}^* \leftarrow C_{\text{total}}, \{q_k^*\} \leftarrow \{q_k\}$
        \ENDIF
    \ENDIF
\ENDFOR
\RETURN Optimal compression ratios $\{q_k^*\}$
\end{algorithmic}
\end{algorithm}

\subsection{Cut Layer Selection Algorithm with Integrated Compression}
Given satellite communication conditions, task requirements, LLM size, and other factors, the split points are determined by the algorithm described below.
\subsubsection{Solution Strategy}
To efficiently solve the joint optimization of LLM splitting and compression configuration, an A*-based alternating optimization algorithm is designed that updates the optimal compression ratio at each stage during the path search.
The essential components of the algorithm are defined as follows:
\begin{itemize}
  \item Startup cost: the accumulated computation and communication delay along the current path from the source node \( (0, 0) \) to \( v \). It reflects the total startup latency before pipelined execution enters steady state:
  \begin{equation}
  g(v) = \sum_{e \in \mathcal{P}_{0 \to v}} C(e),
  \end{equation}
  where \( \mathcal{P}_{0 \to v} \) denotes the partial path ending at node \( v \).

  \item Steady-state cost: the maximum stage-wise effective delay encountered along the partial path, used to estimate pipeline bottleneck:
  \begin{equation}
  \theta(v) = \max_{e \in \mathcal{P}_{0 \to v}} T^{\text{eff}}(e),
  \end{equation}
  where \( T^{\text{eff}}(e) \) is defined as in Eq.~\eqref{T_eff}.

  \item Heuristic estimate: a lower-bound estimate of the remaining delay to reach the terminal node \( (L, K) \). To ensure admissibility, we assume all remaining layers are assigned to the fastest available satellite:
  \begin{equation}
  h(v) = \frac{\sum_{i = l+1}^{L} c_i}{f_{\max}},
  \end{equation}
  where \( c_i \) is the computation cost of layer \( i \), and \( f_{\max} = \max_{j > k} f_j \) is the maximum speed among remaining satellites.
\end{itemize}
The total search cost is formulated as
\begin{equation}
f(v) = g(v) + (B - 1)\theta(v) + h(v),
\end{equation}
which reflects the total inference delay for $B$ pipelined inputs, combining startup cost, steady-state cost, and a heuristic estimate of future cost.

During the search, for each valid layer split from $(l, k)$ to $(l', k+1)$, the algorithm temporarily assigns layers $l+1$ to $l'$ to the $(k+1)$-th satellite. A per-edge compression subproblem is solved (Eq.~\ref{eq:compression_subproblem}) to determine the optimal compression ratio $q_{k+1}$. The resulting $T_{k+1}^{\text{eff}}$ is used to update $g(v)$ and $\theta(v)$. If the new path yields a lower total cost $f(v)$, it is inserted into the priority queue.
The search terminates once the goal node $(L, K)$ is reached, returning the optimal partition $\{l_k\}$ and corresponding compression ratios $\{q_k\}$ that minimize the total pipeline inference delay. The complete procedure is outlined in Alg.~\ref{alg:a_star_joint}.

\subsubsection{Computational Complexity}
Let $L$ be the total number of model layers, $K$ the number of satellites, and $N$ the number of discrete grid values used for each compression ratio $q_k \in \{0, \frac{1}{N}, \dots, 1\}$.
The algorithm explores paths in a layered graph with at most $\binom{L}{K-1}$ valid layer partitions. For each partition candidate, it solves a nested compression subproblem using grid search over $(N+1)^{K-1}$ configurations (as analyzed in Algorithm~\ref{alg:compression_search}).
Thus, the worst-case time complexity of Algorithm~\ref{alg:a_star_joint} is $\mathcal{O}\left( \binom{L}{K-1}  (N+1)^{K-1}  K \right)$, where $\binom{L}{K-1}$ represents the number of ways to divide $L$ layers into $K$ contiguous segments, $(N+1)^{K-1}$ is the number of candidate compression configurations, and $K$ accounts for evaluating all stage-wise delays and constraints per candidate.

\begin{algorithm}[t]
\footnotesize
\caption{\textcolor{black}{DAG-Based Model Splitting Algorithm}}
\label{alg:a_star_joint}
\begin{algorithmic}[1]
\REQUIRE Total layers $L$, satellites $K$, memory budget $M_k^{\text{max}}$
\ENSURE Optimal layer partition $\{l_k\}$ and compression ratios $\{q_k\}$;

\STATE Initialize priority queue $\mathcal{Q}$ with start node $(0,0)$ and empty $\{l_k\}, \{q_k\}$
\WHILE{$\mathcal{Q}$ is not empty}
    \STATE Pop node $(l, k)$ with lowest cost $f = g + h$;
    \IF{$(l = L)$ \AND $(k = K)$}
        \STATE \textbf{return} current path $\{l_k\}$ and corresponding $\{q_k\}$;
    \ENDIF
    \FOR{each valid layer split $l' > l$ for satellite $k+1$}
        \IF{Memory constraint on layers $l+1$ to $l'$ at $S_{k+1}$ violated}
            \STATE \textbf{continue}
        \ENDIF
        \STATE Tentatively assign layers $l+1$ to $l'$ to satellite $S_{k+1}$;
        \STATE Solve compression subproblem (Eq.~\ref{eq:compression_subproblem}) to get optimal $\{q_j\}_{j=1}^{k+1}$ for current $\{l_j\}_{j=1}^{k+1}$;
        \STATE Compute $T_k^{\text{eff}}(l_k, q_k)$ for all previous stages;
        \STATE Update cost-to-come $g' = \sum_{j=1}^{k+1} \left( \frac{C_j(l_j)}{f_j} + \frac{q_j S_j}{r_{\text{sat}}} \right)$;
        \STATE Update max-stage delay $\theta = \max_{j \leq k+1} T_j^{\text{eff}}(l_j, q_j)$;
        \STATE Compute total cost $f' = g' + (B-1) \theta + h$;
        \STATE Push node $(l', k+1)$ with updated $\{l_j\}, \{q_j\}, g', \theta$ into $\mathcal{Q}$;
    \ENDFOR
\ENDWHILE
\STATE \textbf{return} Infeasible
\end{algorithmic}
\end{algorithm}

\section{Performance Evaluation}
\subsection{Experimental Settings}

\subsubsection{Satellite Networks}
In the simulation, a single orbital plane is considered from a Walker Delta constellation with parameters $(1, 12/0, 53^\circ)$, where twelve satellites are evenly distributed in a circular LEO with an altitude of 500\,km and an inclination of $53^\circ$. Each satellite moves along the same orbital trajectory, with an observation window of 10 minutes per cycle. The total simulation period spans 144 time slots, corresponding to a full 24-hour cycle, during which the orbital plane completes one full rotation around the Earth. The observation target and the ground station are located at $(0^\circ\text{N},\,0^\circ\text{E})$ and $(-53^\circ\text{N},\,180^\circ\text{W})$, respectively.
The first satellite in the orbital plane is equipped with an imagery sensor and responsible for data collection. Within each 10-minute time slot, it captures approximately 300 high-resolution remote sensing images, including resolutions of 1080p, 2K, 4K, 8K, and 16K, to simulate diverse data volume and quality demands in practical applications.
The S2G communication utilizes the Ka-band, and the transmission rate is assumed to remain constant within each 15-second interval. The ISLs that offer a stable and high-throughput channel, assuming negligible relative motion between satellites on the same orbital path.
Each computing-enabled satellite is equipped with onboard processing units and a memory capacity of 8\,GB to support edge computing tasks, such as intermediate feature extraction, compression, or partial inference during the data relay and collaboration process.

\subsubsection{Computation Configuration}
The experimental evaluation is based on measurement data collected from a prototype satellite--ground collaborative inference testbed. The ground station is represented by a server equipped with an NVIDIA RTX 4070ti GPU. Satellite-side computation is emulated by four NVIDIA Jetson AGX Orin devices, each capable of delivering up to 275 TOPs of AI performance, which are widely used in practical aerospace systems, such as those deployed on SpaceX satellites~\cite{spaceX}.
The software stack is built on PyTorch for model inference, with FastAPI over HTTP enabling communication. A multi-threaded execution pipeline is implemented to decouple transmission and inference, allowing data processing and transmission to proceed in parallel across nodes.
To simulate the heterogeneous computational capabilities of satellites in orbit, the Jetson AGX Orin devices are configured to operate under different power modes. Specifically, the power limits are set to 15W, 30W, and 50W, corresponding to satellites with heavy computational loads or energy constraints, moderate processing availability, and idle nodes with full capacity, respectively.

\subsubsection{Communication Configuration}
Both satellite-to-ground and inter-satellite communications are considered under the following communication settings.
For the satellite-to-ground link, a carrier frequency of 40\,GHz with a bandwidth of 1\,GHz is considered. 
The satellite transmit power is set to 35\,dBm, and the antenna gain is 37\,dBi. The path loss exponent is configured as 2.5 to account for free-space attenuation. The communication availability is determined by a visibility angle threshold of $50^\circ$.
For the inter-satellite free-space optical link, the laser wavelength is set to 1550\,nm, with a transmit power of 10\,dBW (10\,W). The optical beam has a divergence angle of 50\,$\mu$rad, and the receiver aperture diameter is 0.10\,m. A total optical system loss of 6\,dB is assumed, capturing pointing errors and hardware losses. Thermal noise is modeled using a Boltzmann constant of $1.38 \times 10^{-23}$\,J/K and a system temperature of 290\,K. The available FSO communication bandwidth is set to 0.5\,GHz.

\begin{table}[t]
\centering
\caption{System Parameters in Simulation}
\label{tab:system-params}
\begin{tabular}{l|l}
\toprule
\textbf{Parameter} & \textbf{Value } \\
\hline
Number of satellites & 5 \\
Batch size & 64 \\
Ground server GPU & NVIDIA RTX 4070ti \\
Satellite devices & 4× NVIDIA Jetson AGX Orin \\
Intersatellite link rate & 0.5\,Gbit/s \\
Satellite-ground link rate & 6\,Gbit/s \\
Downlink window & 10 minutes per satellite \\
\bottomrule
\end{tabular}
\end{table}

\begin{table}[t]
\centering
\caption{Model and Dataset Settings}
\label{tab:model-dataset}
\begin{tabular}{l|l}
\toprule
\textbf{Item} & \textbf{Value / Description} \\
\hline
ViT-B & 0.086 Billion parameters, ~2 GB memory \\
ViT-L & 0.307 Billion parameters, ~4 GB memory \\
ViT-H & 0.632 Billion parameters, ~7 GB memory \\
ViT-G & 1.8 Billion parameters, ~12 GB memory \\
EuroSAT & 27,000 RGB images, 10 scene classes \\
RESISC45 & 31,500 RGB images, 45 scene classes \\
\bottomrule
\end{tabular}
\end{table}

\subsubsection{Application Scenario}
A typical satellite-based computer vision application is considered, in which real-time images captured by a remote-sensing satellite are processed collaboratively across a satellite constellation. In particular, an object detection task is considered, and the end-to-end delay from the start of image processing to the delivery of results to the ground is evaluated.
To assess the performance of our collaborative inference scheme across different computational loads, four Vision Transformer (ViT) models of varying sizes are adopted: ViT-B (Base), ViT-L (Large), ViT-H (Huge), and ViT-G (Giant)~\cite{Model1,Model2}. These models differ significantly in both parameter count and memory footprint. Specifically, ViT-B contains approximately 0.086 billion parameters and typically requires around 2 GB of memory during inference. ViT-L increases the parameter count to 0.307 billion and consumes about 4 GB of memory. ViT-H includes 0.632 billion parameters and requires approximately 7 GB of memory. ViT-G, the largest model in our study, has 1.8 billion parameters and demands over 12 GB of memory.

\subsubsection{Datasets}
The EuroSAT dataset is a remote sensing scene classification benchmark based on Sentinel-2 satellite imagery~\cite{Dataset1}. It consists of 10 land use and land cover categories, such as agricultural, forest, and residential areas. The dataset contains a total of 27,000 RGB images with a resolution of 64×64 pixels, divided into 19,500 training images and 7,500 testing images.
The RESISC45 dataset is another scene classification dataset comprising 45 categories, including airport, harbor, desert, stadium, and more~\cite{Dataset2}. It contains 31,500 RGB images with a resolution of 256×256 pixels, with each category represented by 700 images. The dataset is split into 25,200 training images and 6,300 testing images.

\begin{figure}[t]
\centering
\includegraphics[width=0.68\linewidth]{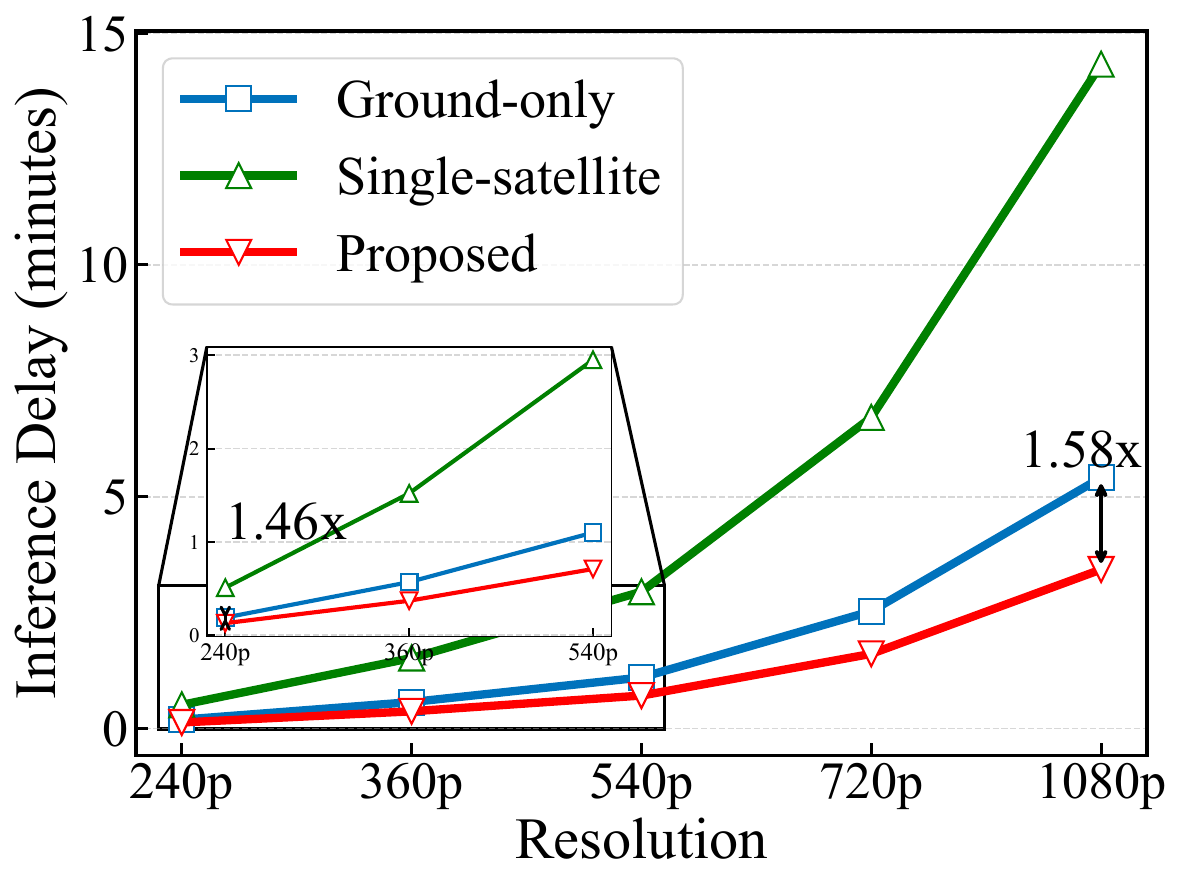}
\caption{Inference delay under varying image resolutions.}
\label{fig_vary_resolution}
\end{figure}

\begin{figure}[t]
\centering
\includegraphics[width=0.68\linewidth]{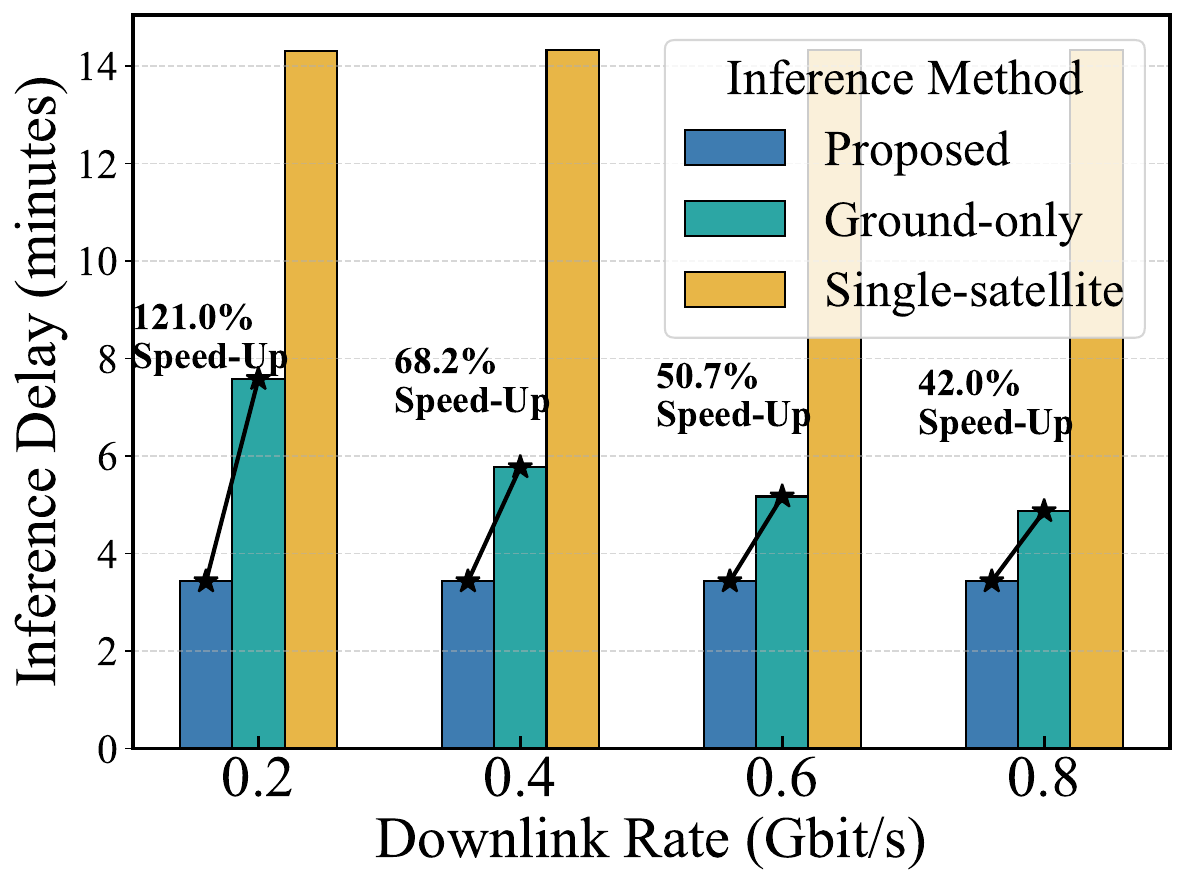}
\caption{Inference delay with respect to S2G rates.}
\label{fig_vary_data}
\end{figure}

\begin{figure}[t]
\centering
\includegraphics[width=0.68\linewidth]{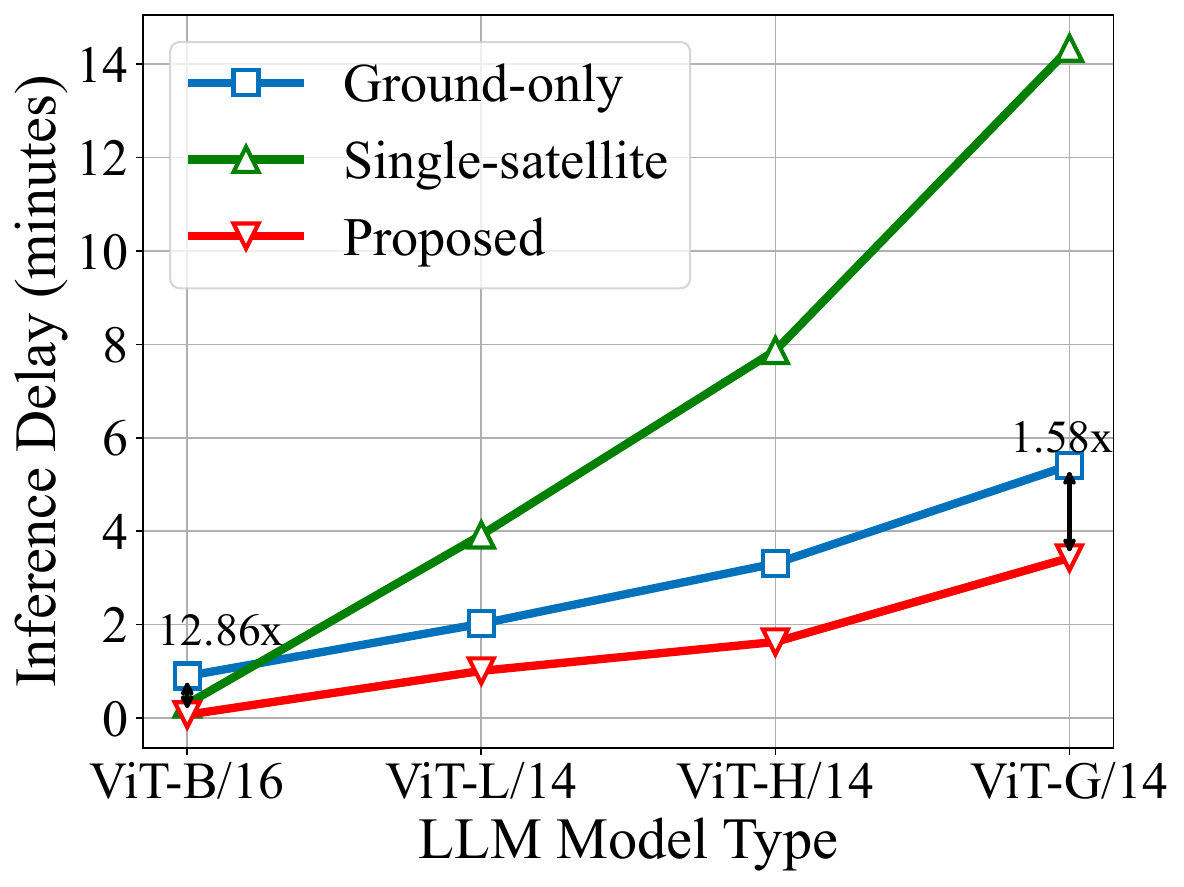}
\caption{Inference delay under different satellite numbers. }
\label{fig_model_complexity}
\end{figure}

\subsubsection{Benchmark Schemes}
To evaluate the performance of our satellite-based collaborative inference framework, the following baseline strategies are considered:
\begin{itemize}
    \item \textbf{Ground-only:} The sensing satellite captures raw images and transmits them to the ground station via $K$ relay satellites. The transmission follows a pipeline-parallel manner, where multiple inputs are continuously forwarded stage by stage. Raw images are downlinked for full-model inference on powerful terrestrial servers. 
    
    \item \textbf{Single-satellite:} The sensing satellite forwards the captured image to a designated computing satellite. The full model is deployed on this satellite to perform the complete inference locally. The final prediction result is then transmitted back to the ground station through relay satellites.
\end{itemize}

\subsubsection{Metrics}The effectiveness of the collaborative inference scheme is evaluated using the following metrics:
\begin{itemize}
    \item \textbf{Inference Latency:} The time interval between the task initiation at the head satellite and the receipt of the final prediction result at the ground station.
    
    \item \textbf{Inference Accuracy:} The classification performance of the model after splitting and compressing intermediate activation.
    
    \item \textbf{Communication Overhead:} The total amount of data transmitted throughout the task execution, including both inter-satellite and satellite-to-ground communication.
    
    \item \textbf{Optimization Gain:} The performance improvement achieved by applying the A*-based graph optimization algorithm for LLM layer assignment to devices.
\end{itemize}


\subsection{Evaluation Results} 
\subsubsection{Inference Delay Performance}
As shown in Fig.~\ref{fig_vary_resolution}, the inference delay of all three schemes increases with data resolution. The proposed scheme consistently achieves the lowest delay, regardless of the input resolution. At 240p, the scheme reduces the delay by at least 46\% compared to the others, and achieves a 58\% reduction at 1080p. The stable gain arises from the fact that higher-resolution images increase both transmission and computation costs. While ground-only inference suffers from heavy downlink burdens, onboard inference is also slowed due to the larger number of split mini-batches required for high-resolution images.

The impact of the S2G transmission rate on inference delay is shown in Fig.~\ref{fig_vary_data}, under four typical downlink configurations within the Ka-band: 0.2, 0.4, 0.6, and 0.8 Gbit/s. As shown in the figure, the proposed collaborative inference scheme consistently achieves the lowest delay across all transmission settings. Notably, the relative improvement diminishes with higher S2G rates, as the ground transmission bottleneck is gradually alleviated. Nevertheless, even at 800 Mbit/s, the proposed scheme still yields a 42.0\% reduction in end-to-end delay compared to the ground-only baseline, and over 100\% improvement in low-rate scenarios.

\begin{figure}[t]
\centering
\includegraphics[width=0.68\linewidth]{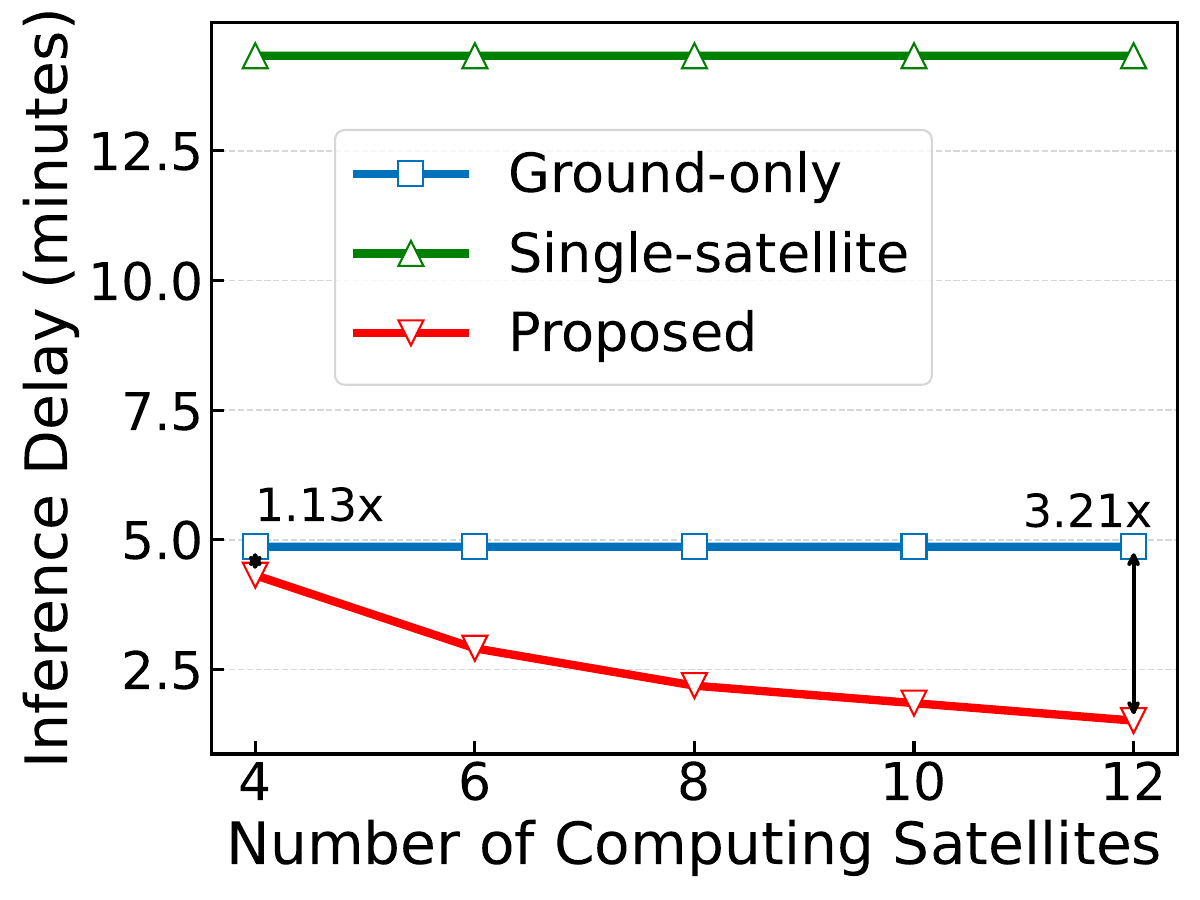}
\caption{Inference delay comparison under varying deployment conditions.}
\label{fig_number}
\end{figure}

\begin{figure}[t]
\centering
\includegraphics[width=0.68\linewidth]{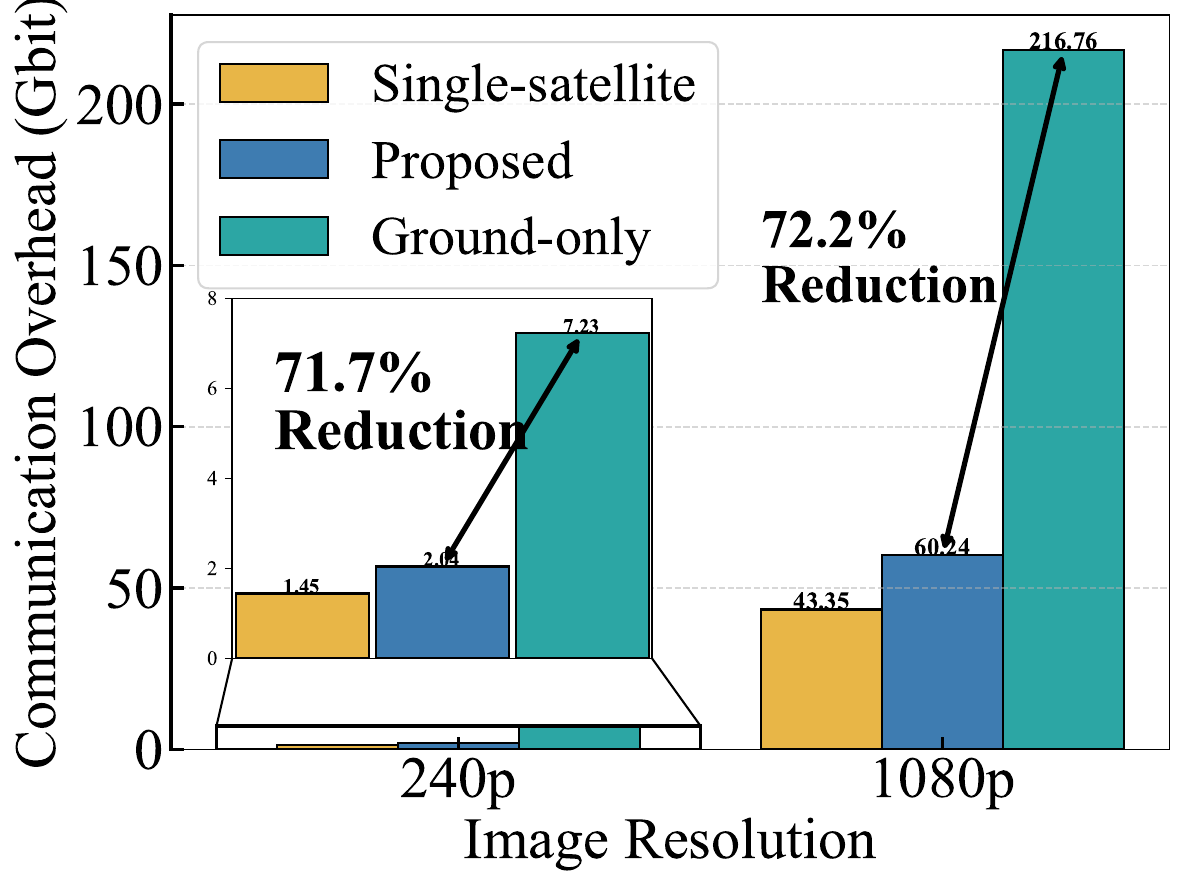}
\caption{Total communication overhead across different schemes.}
\label{fig_communication_overhead}
\end{figure}

\begin{figure}[t]
\centering
\includegraphics[width=0.68\linewidth]{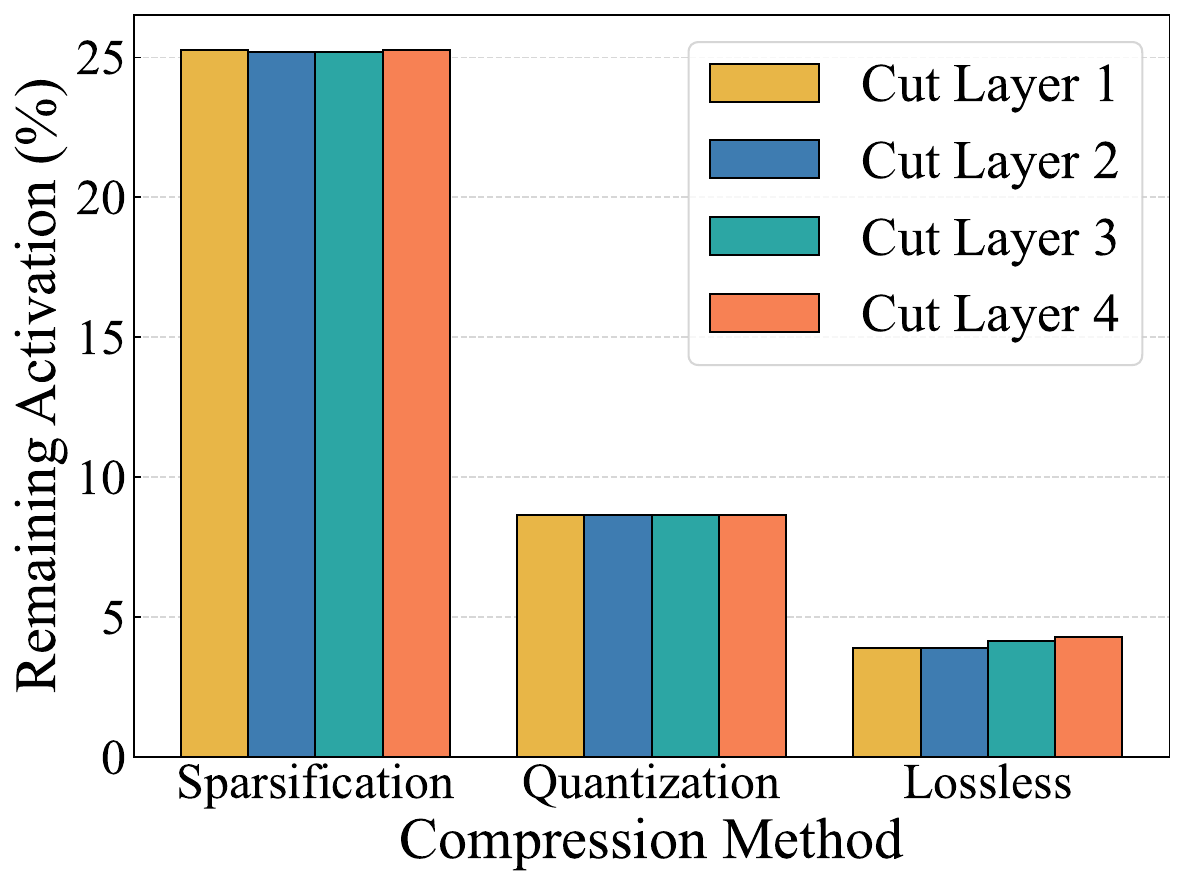}
\caption{Ablation Study of Compression Methods.}
\label{fig_communication_Ablation}
\end{figure}

Figure~\ref{fig_model_complexity} shows the inference delay under different model scales for three deployment schemes. As the model size increases from ViT-B to ViT-g, the performance gain of the proposed scheme gradually diminishes. This is because of the limited onboard computing capacity, where deploying large-scale models induces a considerable computational burden, thus increasing the overall inference delay. For smaller models, the single-satellite scheme demonstrates a stronger advantage, as LLM splitting and multi-satellite collaboration introduce unnecessary overhead. Thus, with five satellites, our scheme is most effective for billion-scale models.

As the number of satellites in a constellation is typically predetermined, it is not feasible to control the number of satellites that are physically deployed. Nonetheless, we conduct a reference analysis to investigate the performance impact of varying the number of satellites participating in the computation, as depicted in Fig.~\ref{fig_number}. The proposed scheme achieves considerable reductions in inference delay as the number of computing satellites increases. The improvement is primarily attributed to the efficient overlap of communication and computation, which becomes more effective as the number of computing nodes increases. However, such performance gains are accompanied by increased inter-satellite communication overhead.

\subsubsection{Communication Overhead}
 As shown in Fig.~\ref{fig_communication_overhead}, the communication overhead required to complete the inference task is compared under both low and high resolution settings. The results demonstrate that the proposed scheme consistently achieves lower communication overhead than the ground-based computation scheme, primarily due to the adoption of intermediate activation compression. Although the single-satellite computation exhibits the lowest communication overhead, it is frequently associated with the highest end-to-end inference delay across most cases. Additionally, the communication gain of the proposed scheme remains stable across different resolutions. This is because, while higher resolution increases the amount of data transmitted, it also raises the number of mini-batches, thereby proportionally increasing the communication overhead for both the proposed and ground-based schemes.

An ablation study on the compression schemes is conducted, as illustrated in Fig.~\ref{fig_communication_Ablation}. For Satellite 1, the activations are successively compressed by the Gumbel-mask-based sparsification, quantization, and lossless encoding, achieving compression ratios of $3.96\times$, $11.56\times$, and $25.82\times$, respectively. For Satellites 2, 3, and 4, the compression ratios of sparsification and quantization remain nearly identical to those of Satellite 1, indicating stable structural compression across layers. However, the effectiveness of lossless encoding decreases to $25.60\times$, $24.14\times$, and $23.28\times$, respectively. This degradation stems from the increasing complexity and dispersion of activation distributions in deeper layers, which leads to higher entropy and thus reduces the efficiency of entropy-based compression.
Nevertheless, it is noteworthy that all activation compression paths consistently achieve over $20\times$ overall compression, ensuring high transmission efficiency.

\begin{figure}[t]
\centering
\includegraphics[width=0.68\linewidth]{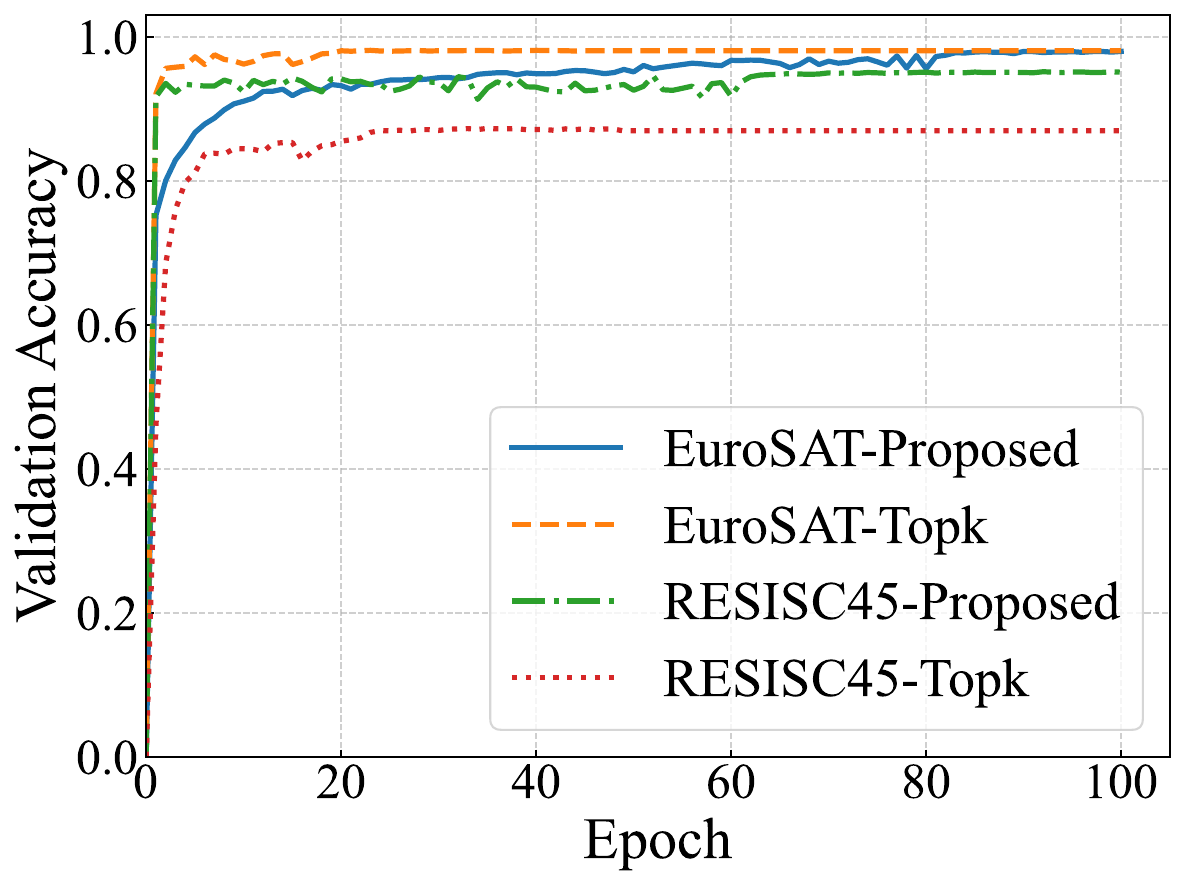}
\caption{Accuracy performance of LLM on the EuroSAT and Resisc45 dataset varying with each training round.}
\label{fig_training_plot}
\end{figure}

\subsubsection{Inference Accuracy} 
As shown in Fig.~\ref{fig_training_plot}, the accuracy curves as a function of the number of epochs are presented under the conditions of the ViT-G model with the EuroSAT and RESISC datasets. The experimental results demonstrate that the proposed Gumbel-mask-based activation compression scheme converges effectively. Furthermore, we conclude that our proposed LLM inference scheme achieves either better or comparable accuracy compared to the Top-K compression scheme, which will be further discussed in the next experiment.
Regarding convergence speed, our proposed scheme requires more training epochs. This is because the compression module contains learnable parameters that must be trained, resulting in additional training overhead. However, because the training process is conducted offline, it does not affect inference delays on LEO satellites. Therefore, the additional training overhead is deemed acceptable for the overall performance of the scheme.

\begin{table}[t!]
\centering
\caption{Classification accuracy (\%) under different compression schemes and ViT model sizes with EuroSAT dataset.}
\label{tab:accuracy_comparison1}
\begin{tabular}{lcccc}
\toprule
\textbf{Model} & \textbf{Baseline} & \textbf{GumbelMask} & \textbf{Top-k}  \\
\midrule
ViT-Tiny   & 98.27 & 98.00 & 98.00  \\
ViT-Base   & 98.52 & 98.06 & 98.01 \\
ViT-Large  & 98.36 & 98.31 & 98.30  \\
\bottomrule
\end{tabular}
\end{table}

\begin{table}[t!]
\centering
\caption{Classification accuracy (\%) under different compression schemes and ViT model sizes with Resisc45 dataset.}
\label{tab:accuracy_comparison2}
\begin{tabular}{lcccc}
\toprule
\textbf{Model} & \textbf{Baseline} & \textbf{GumbelMask} & \textbf{Top-k}  \\
\midrule
ViT-Tiny   & 93.17 & 92.21 & 88.22\\
ViT-Base   & 96.00 & 95.75 & 94.37   \\
ViT-Large  & 96.67 & 95.97 & 95.54  \\

\bottomrule
\end{tabular}
\end{table}

Tables~\ref{tab:accuracy_comparison1} and~\ref{tab:accuracy_comparison2} present the classification accuracy under different compression methods using an 80\% sparsity ratio and 8-bit quantization, across various ViT model sizes on the EuroSAT and Resisc45 datasets, respectively. The proposed GumbelMask consistently achieves accuracy that is close to the baseline and significantly outperforms the Top-k strategy in all settings.
Compared with Top-k, GumbelMask introduces only a minor loss in accuracy, remaining within 1\% of the baseline across different model sizes and datasets. This indicates strong robustness under aggressive compression settings. 
By contrast, Top-k exhibits substantial accuracy degradation, particularly on smaller models. For example, on the EuroSAT dataset (Table II), Top-k causes a drop of about 4.1\% for ViT-Tiny, and the degradation increases to nearly 5\% on the Resisc45 dataset (Table III). This performance loss is attributed to Top-k’s fixed token selection strategy, which lacks adaptability to image-specific feature distributions. As a result, it often removes important tokens that are essential for correct classification, especially in smaller models where each token contributes more critically to the output. In comparison, GumbelMask employs a learned, stochastic selection mechanism that dynamically identifies informative tokens, preserving task-relevant content even under high sparsity.

\begin{figure}[t]
\centering
\includegraphics[width=0.68\linewidth]{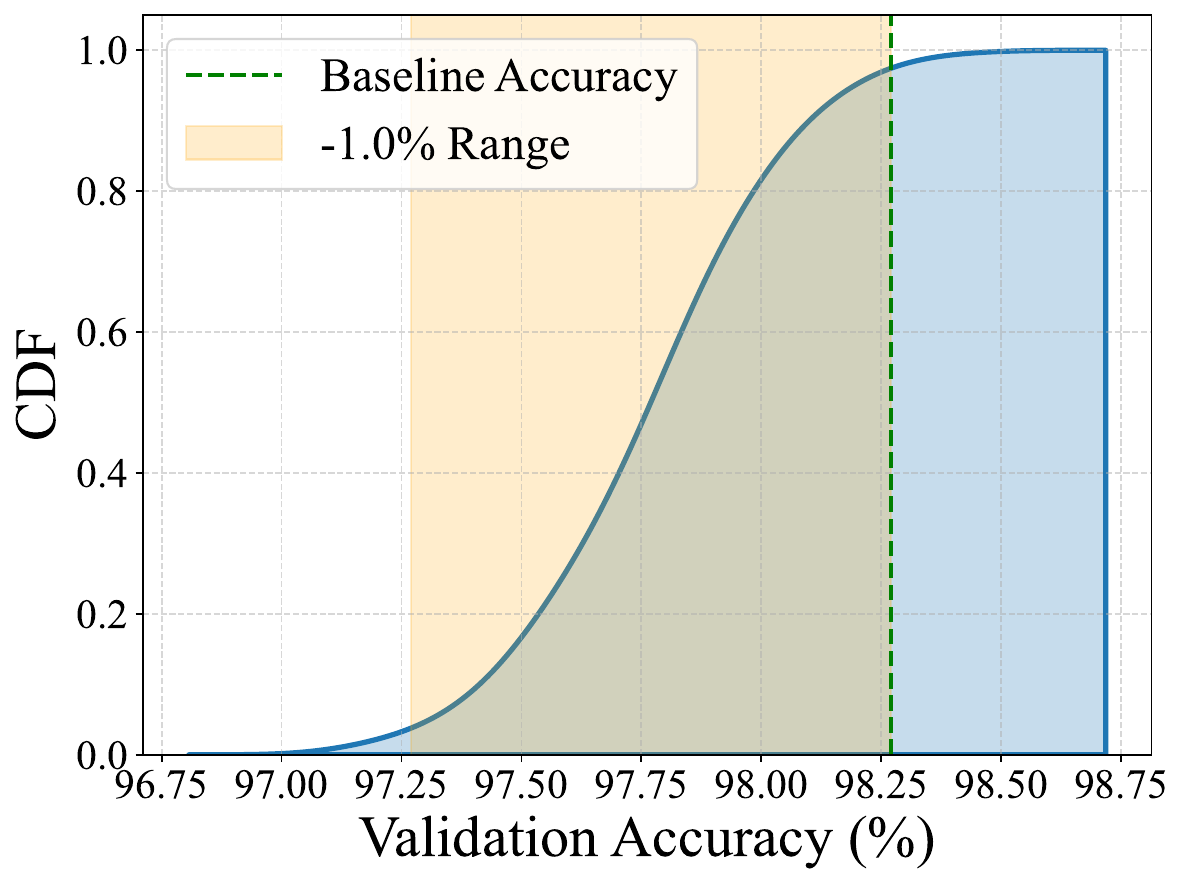}
\caption{ Validation accuracy under various LLM splitting strategies.}
\label{fig_validation}
\end{figure}

Figure~\ref{fig_validation} shows the probability density function of validation accuracy obtained under 200 different LLM splitting strategies. Each splitting strategy corresponds to placing the model division point at a different layer while maintaining a fixed compression configuration. The green dashed line denotes the baseline accuracy without any compression, and the orange shaded region represents a ${-}$1.0\% range around the baseline.
As shown, more than 194 out of 200 validation accuracy measurements fall within a ${-}$1.0\% deviation, indicating stable performance across diverse split points. Only about 6 cases show slightly larger drops, corresponding to deviations of 1.0-1.5\%. These results demonstrate that the proposed compression scheme is insensitive to the location of the model split, ensuring consistent performance regardless of how the model is partitioned.
The robustness is primarily attributable to the learnable nature of the compression scheme, which dynamically adapts to varying split positions and preserves task-relevant information via adaptive activation selection.

\begin{figure}[t]
\centering
\includegraphics[width=0.68\linewidth]{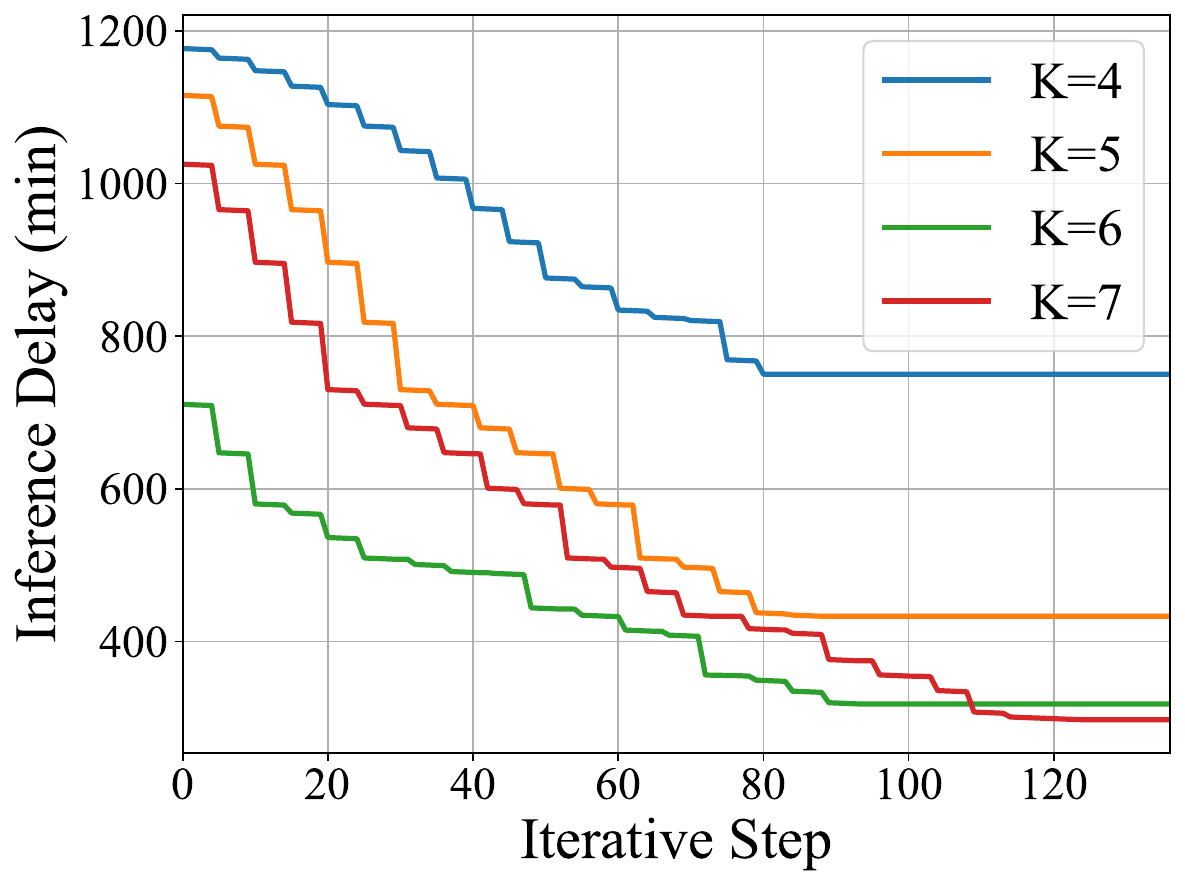}
\caption{Convergence of the proposed algorithm for different satellite numbers.}
\label{fig_convergence}
\end{figure}

\begin{figure}[t]
\centering
\includegraphics[width=0.68\linewidth]{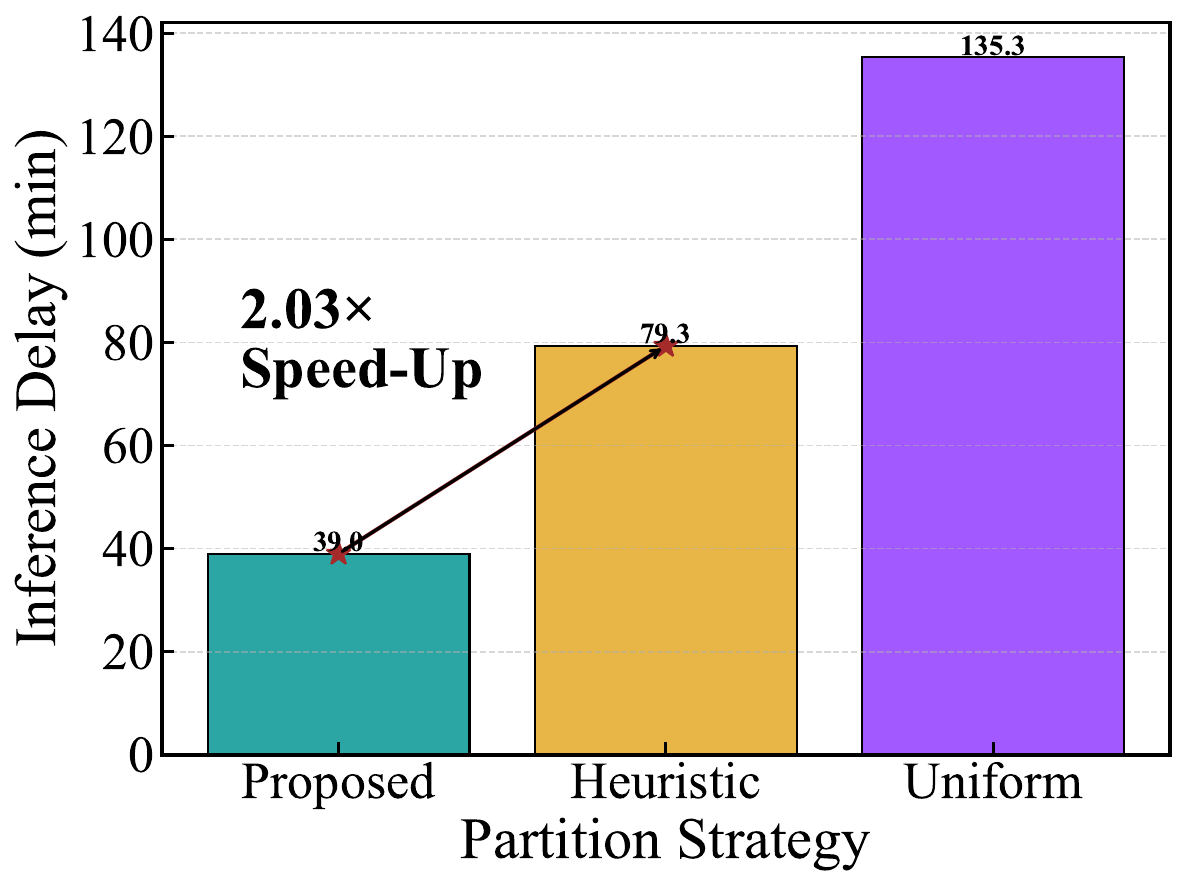}
\caption{Total inference delay under different split strategies.}
\label{fig_optimization}
\end{figure}

\subsubsection{Optimization Gain}
Figure~\ref{fig_convergence} shows the convergence behavior of the proposed joint optimization algorithm.
The experimental results demonstrate that the algorithm consistently converges under different numbers of satellites. As expected, the final inference delay of the algorithm decreases with increasing satellite count. The reason is that more cooperative computing satellites provide greater parallel computational capacity, thus reducing the overall inference delay. Meanwhile, the number of steps required for convergence increases with the addition of more satellites, due to the enlarged search space. Nevertheless, in all tested cases, convergence is achieved within 120 iterative steps.

The inference delays under different LLM splitting strategies are compared in Fig.~\ref{fig_optimization}. The heuristic strategy assigns layers to satellites in proportion to their computational capacity, whereas the uniform strategy divides the layers evenly across all satellites. In the experiment, a 48-layer ViT-Giant model is assigned to five satellites with heterogeneous computing capabilities, emulated by configuring Jetson Orin devices with varying power modes. 
The proposed algorithm achieves a 103\% reduction in total inference delay relative to the heuristic baseline, demonstrating its effectiveness in accelerating inference under heterogeneous resource environments.

\section{Conclusion}\label{8}
In this paper, we have proposed a communication-efficient and collaborative inference scheme by spliting LLM across multiple LEO satellites to enable on-orbit processing. The scheme adopts a pipelined execution strategy that overlaps inter-satellite communication with local computation, thereby accelerating the LLM inference process. 
\textcolor{black}{We have further developed a modified $\mathrm{A}^{\star}$-based search algorithm to optimize the LLM splitting strategy and compression ratios.} 
Experimental results have demonstrated that the proposed scheme outperforms state-of-the-art benchmarks in terms of inference delay, accuracy, and communication overhead, facilitating the practical deployment of computationally intensive LLM tasks in resource-limited satellite networks. For future work, we will investigate collaborative inference schemes across multi-orbit satellite constellations.

\bibliographystyle{IEEEtran}

\bibliography{ref.bib}

@misc{spaceX,
  title        = {SpaceX to launch 1st space-hardened {Nvidia} {AI} {GPU} on upcoming rideshare mission},
  url          = {https://www.space.com/ai-nvidia-gpu-spacex-launch-transporter-11},
  note         = {Accessed: 2024-08-14},
  howpublished = {\url{https://www.space.com/ai-nvidia-gpu-spacex-launch-transporter-11}}
}

@inproceedings{TopK1,
author = {Zheng, Fei and Chen, Chaochao and Lyu, Lingjuan and Yao, Binhui},
title = {Reducing communication for split learning by randomized top-k sparsification},
year = {2023},
booktitle = {Proc. ACM IJCAI},
volume={},
number={519},
pages={4665-4673},
}

@ARTICLE{shen2,
  author={Shen, Xuemin and Gao, Jie and Wu, Wen and Li, Mushu and Zhou, Conghao and Zhuang, Weihua},
  journal={IEEE Commun. Surveys Tuts.}, 
  title={Holistic Network Virtualization and Pervasive Network Intelligence for {6G}}, 
  year={2022},
  volume={24},
  number={1},
  pages={1-30},
  doi={10.1109/COMST.2021.3135829}}

@ARTICLE{shen3,
  author={Wu, Wen and Li, Mushu and Qu, Kaige and Zhou, Conghao and Shen, Xuemin and Zhuang, Weihua and Li, Xu and Shi, Weisen},
  journal={IEEE J. Sel. Areas Commun.}, 
  title={Split Learning Over Wireless Networks: Parallel Design and Resource Management}, 
  year={2023},
  volume={41},
  number={4},
  pages={1051-1066},
  doi={10.1109/JSAC.2023.3242704}}

@ARTICLE{Backgroud1,
  author={Zhang, Ruichen and Du, Hongyang and Liu, Yinqiu and Niyato, Dusit and Kang, Jiawen and Xiong, Zehui and Jamalipour, Abbas and In Kim, Dong},
  journal={IEEE J. Sel. Areas Commun.}, 
  title={Generative {AI} Agents With Large Language Model for Satellite Networks via a Mixture of Experts Transmission}, 
  year={2024},
  volume={42},
  number={12},
  pages={3581-3596},
  doi={10.1109/JSAC.2024.3459037}}

@ARTICLE{Background2,
  author={Lu, Anqi and Hu, Youbing and Cao, Zhiqiang and Liu, Jie and Li, Lingzhi and Li, Zhijun},
  journal={IEEE Trans. Mobile Comput.}, 
  title={Enhancing Remote Sensing Image Scene Classification With Satellite-Terrestrial Collaboration and Attention-Aware Transmission Policy}, 
  year={2025},
  volume={24},
  number={5},
  pages={4496-4509},
  doi={10.1109/TMC.2025.3526142}}

@ARTICLE{Background3,
  author={Shi, Yuanming and Zeng, Li and Zhu, Jingyang and Zhou, Yong and Jiang, Chunxiao and Letaief, Khaled B.},
  journal={IEEE Trans. Wireless Commun.}, 
  title={Satellite Federated Edge Learning: Architecture Design and Convergence Analysis}, 
  year={2024},
  volume={23},
  number={10},
  pages={15212-15229},
  doi={10.1109/TWC.2024.3427377}}

@inproceedings{Model2,
  title={Scaling vision transformers to 22 billion parameters},
  author={Dehghani, Mostafa and Djolonga, Josip and Mustafa, Basil and Padlewski, Piotr and Heek, Jonathan and Gilmer, Justin and Steiner, Andreas Peter and Caron, Mathilde and Geirhos, Robert and Alabdulmohsin, Ibrahim and others},
  booktitle={Proc. PMLR},
  pages={7480--7512},
  year={2023},
  organization={PMLR}
}

@inproceedings{Model1,
  title     = {An Image is Worth 16x16 Words: Transformers for Image Recognition at Scale},
  author    = {Alexey Dosovitskiy and Lucas Beyer and Alexander Kolesnikov and Dirk Weissenborn and Xiaohua Zhai and Thomas Unterthiner and Mostafa Dehghani and Matthias Minderer and Georg Heigold and Sylvain Gelly and Jakob Uszkoreit and Neil Houlsby},
  booktitle = {Proc. ICLR},
  pages={1--21},
  year      = {2021},
}

@ARTICLE{YuhaoChen,
  author={Chen, Yuhao and Yang, Qianqian and He, Shibo and Shi, Zhiguo and Chen, Jiming and Guizani, Mohsen},
  journal={IEEE Trans. Mobile Comput.}, 
  title={FTPipeHD: A Fault-Tolerant Pipeline-Parallel Distributed Training Approach for Heterogeneous Edge Devices}, 
  year={2024},
  volume={23},
  number={4},
  pages={3200-3212},
  doi={10.1109/TMC.2023.3272567}}

@inproceedings{Pipedream,
  title={PipeDream: Generalized pipeline parallelism for DNN training},
  author={Narayanan, Deepak and Harlap, Aaron and Phanishayee, Amar and Seshadri, Vivek and Devanur, Nikhil R and Ganger, Gregory R and Gibbons, Phillip B and Zaharia, Matei},
  booktitle={Proc. ACM SOSP},
  pages={1--15},
  year={2019}
}

@ARTICLE{Songge2,
  author    = {Zhang, Songge and Cheng, Guoliang and Wu, Wen and Huang, Xinyu and Song, Lingyang and Shen, Xuemin},
  title     = {Split Fine-Tuning for Large Language Models in Wireless Networks},
  journal   = {IEEE J. Sel. Topics Signal Process.},
  year      = {2025},
  month     = {Jun. 19},
  note      = {{Early} Access, doi:10.1109/JSTSP.2025.3581484},
  doi       = {10.1109/JSTSP.2025.3581484}
}

@ARTICLE{Songge1,
  author    = {Zhang, Songge and Wu, Wen and Song, Lingyang and Shen, Xuemin},
  title     = {Efficient Model Training in Edge Networks with Hierarchical Split Learning},
  journal   = {IEEE Trans. Mobile Comput.},
  year={2025},
  volume={24},
  number={10},
  pages={10214-10229}
}

@ARTICLE{Back1,
  author={Giuffrida, Gianluca and Fanucci, Luca and Meoni, Gabriele and Batič, Matej and Buckley, Léonie and Dunne, Aubrey and van Dijk, Chris and Esposito, Marco and Hefele, John and Vercruyssen, Nathan and Furano, Gianluca and Pastena, Massimiliano and Aschbacher, Josef},
  journal={IEEE Trans. Geosci. Remote Sens.}, 
  title={The {Sat}-1 Mission: The First On-Board Deep Neural Network Demonstrator for Satellite Earth Observation}, 
  year={2022},
  volume={60},
  number={1},
  pages={1-14},
  doi={10.1109/TGRS.2021.3125567}}

@ARTICLE{Back2,
  author={Cratere, Angela and Farissi, M. Salim and Carbone, Andrea and Asciolla, Marcello and Rizzi, Maria and Dell’Olio, Francesco and Nascetti, Augusto and Spiller, Dario},
  journal={IEEE J. Miniatur. Air Space Syst.}, 
  title={Efficient {FPGA}-Accelerated Convolutional Neural Networks for Cloud Detection on CubeSats}, 
  year={2025},
  month     = {Jan. 15},
  note      = {{Early} Access, doi:10.1109/TMC.2025.3569407},
  doi={10.1109/JMASS.2025.3533018}}

@ARTICLE{Back4,
  author={Zhang, Wenhua and Deng, Wenjing and Cui, Zhen and Liu, Jia and Jiao, Licheng},
  journal={IEEE Geosci. Remote Sens. Lett.}, 
  title={Object Knowledge Distillation for Joint Detection and Tracking in Satellite Videos}, 
  year={2024},
  volume={62},
  number={1},
  pages={1-13},
  doi={10.1109/TGRS.2024.3355933}}

@ARTICLE{Back5,
  author={Sun, Jianfei and Wu, Cong and Mumtaz, Shahid and Tao, Junyi and Cao, Mingsheng and Wang, Mei and Frascolla, Valerio},
  journal={IEEE J. Sel. Areas Commun.}, 
  title={An Efficient Privacy-Aware Split Learning Framework for Satellite Communications}, 
  year={2024},
  volume={42},
  number={12},
  pages={3355-3365},
  doi={10.1109/JSAC.2024.3459027}}

@ARTICLE{Back6,
  author={Wu, Jing and Wang, Lin and Pei, Qiangyu and Cui, Xingqi and Liu, Fangming and Yang, Tingting},
  journal={IEEE Trans. Parallel Distrib. Syst.}, 
  title={{HiTDL}: High-Throughput Deep Learning Inference at the Hybrid Mobile Edge}, 
  year={2022},
  volume={33},
  number={12},
  pages={4499-4514},
  doi={10.1109/TPDS.2022.3195664}}

@ARTICLE{Back7,
  author={Li, Jing and Liang, Weifa and Li, Yuchen and Xu, Zichuan and Jia, Xiaohua and Guo, Song},
  journal={IEEE J. Sel. Areas Commun.}, 
  title={Throughput Maximization of Delay-Aware {DNN} Inference in Edge Computing by Exploring {DNN} Model Partitioning and Inference Parallelism}, 
  year={2023},
  volume={22},
  number={5},
  pages={3017-3030},
  doi={10.1109/TMC.2021.3125949}}

@ARTICLE{Back8,
  author={Cao, Xuelin and Yang, Bo and Shen, Yulong and Yuen, Chau and Zhang, Yan and Han, Zhu and Poor, H. Vincent and Hanzo, Lajos},
  journal={IEEE J. Sel. Areas Commun.}, 
  title={Edge-Assisted Multi-Layer Offloading Optimization of {LEO} Satellite-Terrestrial Integrated Networks}, 
  year={2023},
  volume={41},
  number={2},
  pages={381-398},
  doi={10.1109/JSAC.2022.3227032}}

@ARTICLE{Back9,
  author={Xie, Renchao and Tang, Qinqin and Wang, Qiuning and Liu, Xu and Yu, F. Richard and Huang, Tao},
  journal={IEEE Network}, 
  title={Satellite-Terrestrial Integrated Edge Computing Networks: Architecture, Challenges, and Open Issues}, 
  year={2020},
  volume={34},
  number={3},
  pages={224-231},
  doi={10.1109/MNET.011.1900369}}

@article{Xinyu,
  title={{HiveMind}: Towards cellular native machine learning model splitting},
  author={Wang, Song and Zhang, Xinyu and Uchiyama, Hiromasa and Matsuda, Hiroki},
  journal={IEEE J. Sel. Areas Commun.},
  volume={40},
  number={2},
  pages={626--640},
  year={2021},
  publisher={IEEE}
}

@ARTICLE{Dataset1,
  author={Helber, Patrick and Bischke, Benjamin and Dengel, Andreas and Borth, Damian},
  journal={IEEE J. Sel. Topics Appl. Earth Observ. Remote Sens.}, 
  title={{EuroSAT}: A Novel Dataset and Deep Learning Benchmark for Land Use and Land Cover Classification}, 
  year={2019},
  volume={12},
  number={7},
  pages={2217-2226},
  doi={10.1109/JSTARS.2019.2918242}}

@article{Dataset2,
  title={Remote sensing image scene classification: Benchmark and state of the art},
  author={Cheng, Gong and Han, Junwei and Lu, Xiaoqiang},
  journal={Proc. IEEE},
  volume={105},
  number={10},
  pages={1865--1883},
  year={2017},
  publisher={IEEE}
}

@inproceedings{Quantization,
  title={OPTQ: Accurate Quantization for Generative Pre-trained Transformers},
  author={Frantar, Elias and Ashkboos, Saleh and Hoefler, Torsten and Alistarh, Dan},
  booktitle = {Proc. ICLR},
  pages={1--16},
  year      = {2023}
}

@inproceedings{Pruning,
  title={Movement pruning: Adaptive sparsity by fine-tuning},
  author={Sanh, Victor and Wolf, Thomas and Rush, Alexander},
  booktitle = {Proc. NeurIPS},
  pages={20378--20389},
  year={2020}
}

@inproceedings{Distillation,
title = {MINILM: deep self-attention distillation for task-agnostic compression of pre-trained transformers},
author = {Wang, Wenhui and Wei, Furu and Dong, Li and Bao, Hangbo and Yang, Nan and Zhou, Ming},
  booktitle = {Proc. NeurIPS},
  pages={5776-5788},
  year={2020}
}

@ARTICLE{Vincentpoor,
  author={Cao, Xuelin and Yang, Bo and Shen, Yulong and Yuen, Chau and Zhang, Yan and Han, Zhu and Poor, H. Vincent and Hanzo, Lajos},
  journal={IEEE J. Sel. Areas Commun.}, 
  title={Edge-Assisted Multi-Layer Offloading Optimization of LEO Satellite-Terrestrial Integrated Networks}, 
  year={2023},
  volume={41},
  number={2},
  pages={381-398},
  doi={10.1109/JSAC.2022.3227032}}

@INPROCEEDINGS{ShangguangWang,
  author={Guan, Jinglong and Zhang, Qiyang and Murturi, Ilir and Donta, Praveen Kumar and Dustdar, Schahram and Wang, Shangguang},
  booktitle={Proc. IEEE ICC}, 
  title={Collaborative Inference in {DNN}-Based Satellite Systems with Dynamic Task Streams}, 
  year={2024},
  volume={},
  number={},
  pages={3803-3808},
  doi={10.1109/ICC51166.2024.10622590}}

@INPROCEEDINGS{Chenxu,
  author={Wang, Yuanming and Zhao, Kongyange and Zhang, Xiaoxi and Chen, Xu},
  booktitle={Proc. IEEE INFOCOM WKSHPS}, 
  title={Towards Space Intelligence: Adaptive Scheduling of Satellite-Ground Collaborative Model Inference with Space Edge Computing}, 
  year={2024},
  volume={},
  number={},
  pages={1-6},
  doi={10.1109/INFOCOMWKSHPS61880.2024.10620909}}

@ARTICLE{OJVT,
  author={Shen, Xuemin and Gao, Jie and Wu, Wen and Lyu, Kangjia and Li, Mushu and Zhuang, Weihua and Li, Xu and Rao, Jaya},
  journal={IEEE Open J. Veh. Technol}, 
  title={{AI}-Assisted Network-Slicing Based Next-Generation Wireless Networks}, 
  year={2020},
  volume={1},
  number={},
  pages={45-66},
  doi={10.1109/OJVT.2020.2965100}}

@inproceedings{INFOCOM,
  title={Accelerating End-Cloud Collaborative Inference via Near Bubble-Free Pipeline Optimization},
  author={Gao, Luyao and Liu, Jianchun and Xu, Hongli and Xu, Sun and Ma, Qianpiao and Huang, Liusheng},
  booktitle={Proc. IEEE INFOCOM},
  pages={1--10},
  year={2025},
}

@ARTICLE{ShiwenMao,
  author={Xu, Guanyu and Hao, Zhiwei and Luo, Yong and Hu, Han and An, Jianping and Mao, Shiwen},
  journal={IEEE Trans. Mobile Comput.}, 
  title={De{ViT}: Decomposing Vision Transformers for Collaborative Inference in Edge Devices}, 
  year={Sep. 2024},
  volume={23},
  number={5},
  pages={5917-5932},
  doi={10.1109/TMC.2023.3315138}}

@article{XiaopingZhang2,
  title={Dual vision transformer},
  author={Yao, Ting and Li, Yehao and Pan, Yingwei and Wang, Yu and Zhang, XiaoPing and Mei, Tao},
  journal={IEEE Trans. Pattern Anal. Mach. Intell.},
  volume={45},
  number={9},
  pages={10870--10882},
  year={Apr. 2023},
  publisher={IEEE}
}

@inproceedings{NLP,
  title={Efficient federated learning for modern {NLP}},
  author={Cai, Dongqi and Wu, Yaozong and Wang, Shangguang and Lin, Felix Xiaozhu and Xu, Mengwei},
  booktitle={Proc. ACM Mobicom},
  pages={1--16},
  year={Oct. 2023}
}

@ARTICLE{ZhangHongKe,
  author={Yao, Su and Lin, Yiying and Wang, Mu and Xu, Ke and Xu, Mingwei and Xu, Changqiao and Zhang, Hongke},
  journal={IEEE J. Sel. Areas Commun.}, 
  title={{LEOEdge}: A Satellite-Ground Cooperation Platform for the {AI} Inference in Large {LEO} Constellation}, 
  year={2025},
  volume={43},
  number={1},
  pages={36-50},
}

@ARTICLE{LiWu,
  author={Li, Liang and Yang, Xingke and Wu, Wen and Wang, Hao and Ohtsuki, Tomoaki and Fu, Xin and Pan, Miao and Shen, Xuemin},
  journal={IEEE J. Sel. Topics Signal Process.}, 
  title={MobiLLM: Enabling On-Device Fine-Tuning of Billion-Sized LLMs via Server-Assisted Side-Tuning}, 
  year      = {2025},
  month     = {Nov. 17},
  note      = {{Early} Access, doi:10.1109/JSTSP.2025.3633550},}

@ARTICLE{WUWC,
  author={Wu, Wen and Zhou, Conghao and Li, Mushu and Wu, Huaqing and Zhou, Haibo and Zhang, Ning and Shen, Xuemin Sherman and Zhuang, Weihua},
  journal={IEEE Wireless Commun.}, 
  title={{AI}-Native Network Slicing for {6G} Networks}, 
  year={2022},
  volume={29},
  number={1},
  pages={96-103},
  doi={10.1109/MWC.001.2100338}}
\end{document}